\documentclass[traditabstract,longauth,a4paper]{aa} 
\usepackage{graphicx}
\usepackage{txfonts}
\usepackage{hyperref}
\usepackage{geometry}
\usepackage{epstopdf, epsfig}
\usepackage[usenames]{color}

\newcommand\ergcms{erg\,cm$^{-2}$\,s$^{-1}$\xspace}%
\newcommand\cmsmev{cm$^{-2}$\,s$^{-1}$\,MeV$^{-1}$\xspace}%
\newcommand\phcms{ph\,cm$^{-2}$\,s$^{-1}$\xspace}%
\newcommand\g{\ensuremath{\gamma}}%
\newcommand\hess{H.E.S.S.\xspace}
\newcommand\fermi{\textit{Fermi}/LAT\xspace}
\newcommand\swift{\textit{Swift}\xspace}
\newcommand\pks{\object{PKS\,0301$-$243}\xspace}

\begin{document}

   \title{Discovery of very high energy \g-ray emission from the BL\,Lac object \pks with \hess}
   \titlerunning{Discovery of VHE \g-rays from \pks with \hess}

   \author{H.E.S.S. Collaboration
\and A.~Abramowski \inst{1}
\and F.~Acero \inst{2}
\and F.~Aharonian \inst{3,4,5}
\and F.~Ait Benkhali \inst{3}
\and A.G.~Akhperjanian \inst{6,5}
\and E.~Ang\"uner \inst{7}
\and G.~Anton \inst{8}
\and S.~Balenderan \inst{9}
\and A.~Balzer \inst{10,11}
% ...
\vspace*{-0.3cm}
}

\institute{
Universit\"at Hamburg, Institut f\"ur Experimentalphysik, Luruper Chaussee 149, D 22761 Hamburg, Germany \and
Laboratoire Univers et Particules de Montpellier, Universit\'e Montpellier 2, CNRS/IN2P3,  CC 72, Place Eug\`ene Bataillon, F-34095 Montpellier Cedex 5, France \and
Max-Planck-Institut f\"ur Kernphysik, P.O. Box 103980, D 69029 Heidelberg, Germany \and
Dublin Institute for Advanced Studies, 31 Fitzwilliam Place, Dublin 2, Ireland \and
National Academy of Sciences of the Republic of Armenia, Yerevan  \and
Yerevan Physics Institute, 2 Alikhanian Brothers St., 375036 Yerevan, Armenia \and
Institut f\"ur Physik, Humboldt-Universit\"at zu Berlin, Newtonstr. 15, D 12489 Berlin, Germany \and
Universit\"at Erlangen-N\"urnberg, Physikalisches Institut, Erwin-Rommel-Str. 1, D 91058 Erlangen, Germany \and
University of Durham, Department of Physics, South Road, Durham DH1 3LE, U.K. \and
DESY, D-15735 Zeuthen, Germany \and
Institut f\"ur Physik und Astronomie, Universit\"at Potsdam,  Karl-Liebknecht-Strasse 24/25, D 14476 Potsdam, Germany
% ...
}

   \offprints{D.~Wouters, J.-P.~Lenain\\
     \email{\href{mailto:denis.wouters@cea.fr}{denis.wouters@cea.fr}\\
       \href{mailto:jlenain@lpnhe.in2p3.fr}{jlenain@lpnhe.in2p3.fr}}
   }

   \date{Received 4 April 2013 / Accepted DAY MONTH 2013}

   
  \abstract{The active galactic nucleus \pks ($z=0.266$) is a high-synchrotron-peaked BL\,Lac object that is detected at high energies  (HE, 100 MeV $< E <$ 100 GeV) by \fermi. This paper reports on the discovery of \pks at very high energies ($E>100$\,GeV) by the High Energy Stereoscopic System (\hess) from observations between September 2009 and December 2011 for a total live time of 34.9 hours. Gamma rays above 200\,GeV are detected at a significance of 9.4$\sigma$. A hint of variability at the $2.5\sigma$ level is found. An integral flux $I(E > 200\,\mathrm{GeV} ) = (3.3 \pm 1.1_\mathrm{stat} \pm 0.7_\mathrm{syst}) \times 10^{-12}$\,\phcms and a photon index $\Gamma = 4.6 \pm 0.7_\mathrm{stat} \pm 0.2_\mathrm{syst}$ are measured. Multi-wavelength light curves in HE, X-ray and optical bands show strong variability, and a minimal variability timescale of eight days is estimated from the optical light curve. A single-zone leptonic synchrotron self-Compton scenario satisfactorily reproduces the multi-wavelength data. In this model, the emitting region is out of equipartition and the jet is particle dominated. Because of its high redshift compared to other sources observed at TeV energies, the very high energy emission from \pks is attenuated by the extragalactic background light (EBL) and the measured spectrum is used to derive an upper limit on the opacity of the EBL.}

\addtolength{\textheight}{0.3cm}
   \maketitle
\addtolength{\textheight}{-0.3cm}

%

\section{Introduction}

Active galactic nuclei (AGN) detected at very high energies (VHE; $E>100$\,GeV) usually belong to the class of BL\,Lac objects; there are a few exceptions, radio galaxies \citep{2009ApJ...695L..40A,2012ApJ...746..151A,2012A+A...539L...2A} or flat-spectrum radio quasars \citep[FSRQ,][]{2008Sci...320.1752M,2011ApJ...730L...8A,2013A+A...554A.107H}, for example. While FSRQ have broad emission lines \citep{1991ApJS...76..813S}, BL\,Lac objects are characterised by weak lines in the optical band, or even featureless spectra, with their emission dominated at all wavelengths by their relativistic jets; BL\,Lac objects and FSRQ form the class of blazars. Their spectral energy distribution (SED) presents two broad peaks, the first of which is understood as being due to synchrotron radiation at lower energies. The high energy peak is commonly explained in leptonic frameworks as inverse-Compton radiation \citep[see e.g.][]{1965ARA+A...3..297G,1985MNRAS.212..553S,1989ApJ...340..181G}, but hadronic models represent a viable alternative \citep[see e.g.][]{1991A+A...251..723M,2000NewA....5..377A,2001APh....15..121M}. The BL\,Lac objects are split into three additional categories, depending on the frequency of the peak of the synchrotron component. The synchrotron emission of the low-frequency-peaked BL\,Lac objects \citep[LBL, see ][]{1995ApJ...444..567P} typically peaks below $10^{14}$\,Hz, above $10^{15}$\,Hz \citep{1996MNRAS.279..526P} for high-frequency-peaked BL\,Lac objects (HBL), and in between for intermediate-frequency-peaked BL\,Lac objects \citep[IBL, see][]{1998ApJS..118..127L,1999ApJ...525..127L}.

While propagating to the Earth, VHE \g-rays experience absorption by the extragalactic background light \citep[EBL,][]{2001ARA+A..39..249H,2005PhR...409..361K,1992ApJ...390L..49S}, which in turn makes TeV emitting AGN interesting probes to study the EBL independently from other measurements such as galaxy counts \citep[see e.g.][]{2006A+A...451..417D}. Imaging atmospheric Cherenkov telescopes (IACT) put strong constraints on the shape and the density level of the EBL, through studies of distant HBL objects \citep[see e.g.][]{2006Natur.440.1018A,2008Sci...320.1752M,2013A+A...550A...4H}.

The object \pks was first identified as a blazar by \citet{1988ApJ...333..666I} with a high polarimetric fraction in the optical regime. It was first classified as an LBL by \citet{1996A+A...311..384L}, whose classification was revised to intermediate-synchrotron-peaked blazar by \citet{2010ApJ...715..429A} (but see Sect.~\ref{subsec-mwlsed}), but was then reclassified as a high-synchrotron-peaked blazar by \citet{2010ApJ...716...30A}. Based on a spectroscopic measurement of the redshift of a close galaxy (named G2) taken in January 1994 on the New Technology Telescope (NTT) at La Silla, \citet{1995AJ....110.1554P} suggested that \pks could lie at $z \sim 0.26$. This result was supported by further observations taken in January 1996 at the NTT, with the plausible identification of a single weak emission line with [\ion{O}{III}] 5007 \AA\ in the spectrum of \pks \citep{2000A+A...357...91F}. The redshift was refined by \cite{2012AIPC.1505..566P} to a value of 0.266 with an improved spectroscopy using XSHOOTER at the VLT.

At higher energies, \pks was previously detected in the X-rays using the \textit{ROSAT} satellite \citep{1996A+A...311..384L} and emerged as a bright source in the \textit{ROSAT} All-Sky Survey \citep{1999A+A...349..389V}. While no pointed observation with \textit{XMM} exists for this source, \pks has been detected in the first catalogue of \textit{XMM} slew sources \citep[version 1.5,][]{2008A+A...480..611S}, on August 9, 2009, with a flux of $F_{0.2-12\,\mathrm{keV}} = (1.4 \pm 0.4) \times 10^{-12}$\,\ergcms.

The high redshift (for VHE studies) together with the association of \object{0FGL\,J0303.7$-$2410} with \pks in the \textit{Fermi} Bright Source List \citep{2009ApJS..183...46A} motivated observations of \pks with the High Energy Stereoscopic System (H.E.S.S.) to study the imprint of the EBL on the VHE spectra of TeV blazars further. The results of \hess observations between 2009 and 2011 are described in Sect.~\ref{sec-hess}. Data analysis of multi-wavelength data from \fermi, \swift, and ATOM are presented in Sect.~\ref{sec-mwl}. In Sect.~\ref{sec-discussion}, a single-zone leptonic synchrotron self-Compton model is proposed to account for the broadband SED of \pks, and to constrain the radiative mechanisms at work in these sources. The high and very high energy data are used to constrain the opacity of the EBL. These results are summarized in Sect.~\ref{sec-conclusion}. In the following, a $\Lambda$CDM cosmology with $H_0 $= 70\,km\,s$^{-1}$\,Mpc$^{-1}$, $\Omega_\mathrm{m} = 0.27$, and $\Omega_{\Lambda} = 0.73$ is assumed.

\section{\hess observations and results}
\label{sec-hess}

The High Energy Stereoscopic System is an array of four IACTs~\citep{2006A+A...457..899A}, located in the Khomas Highland of Namibia, that is used to observe VHE \g-rays above an energy threshold of $\sim$ 100 GeV.  Some of the main features are an angular resolution of $\sim 0.1 \degr$ and an energy resolution of $\sim$ 15 \%. More details about the \hess experiment are given in \cite{2004NewAR..48..331H}.

The object \pks was been observed between August 2009 and December 2011 for a total observation time of 58.5\,h. After data quality selection and dead-time correction, a total of 34.9\,h of high quality data remains to be used in the analysis. Data are taken at zenith angles ranging from 0\degr\ to 20\degr\ and using the so-called \textit{wobble} mode where a pointing offset from \pks of 0.5\degr\ is maintained in order to simultaneously evaluate the signal and the background from the same field of view. These data are analysed with the \textit{model} analysis \citep{2009APh....32..231D}. The analysis is cross-checked with a multivariate method \citep{2011APh....34..858B}, which yields consistent results.

\begin{figure}
\centering
\includegraphics[width=\columnwidth]{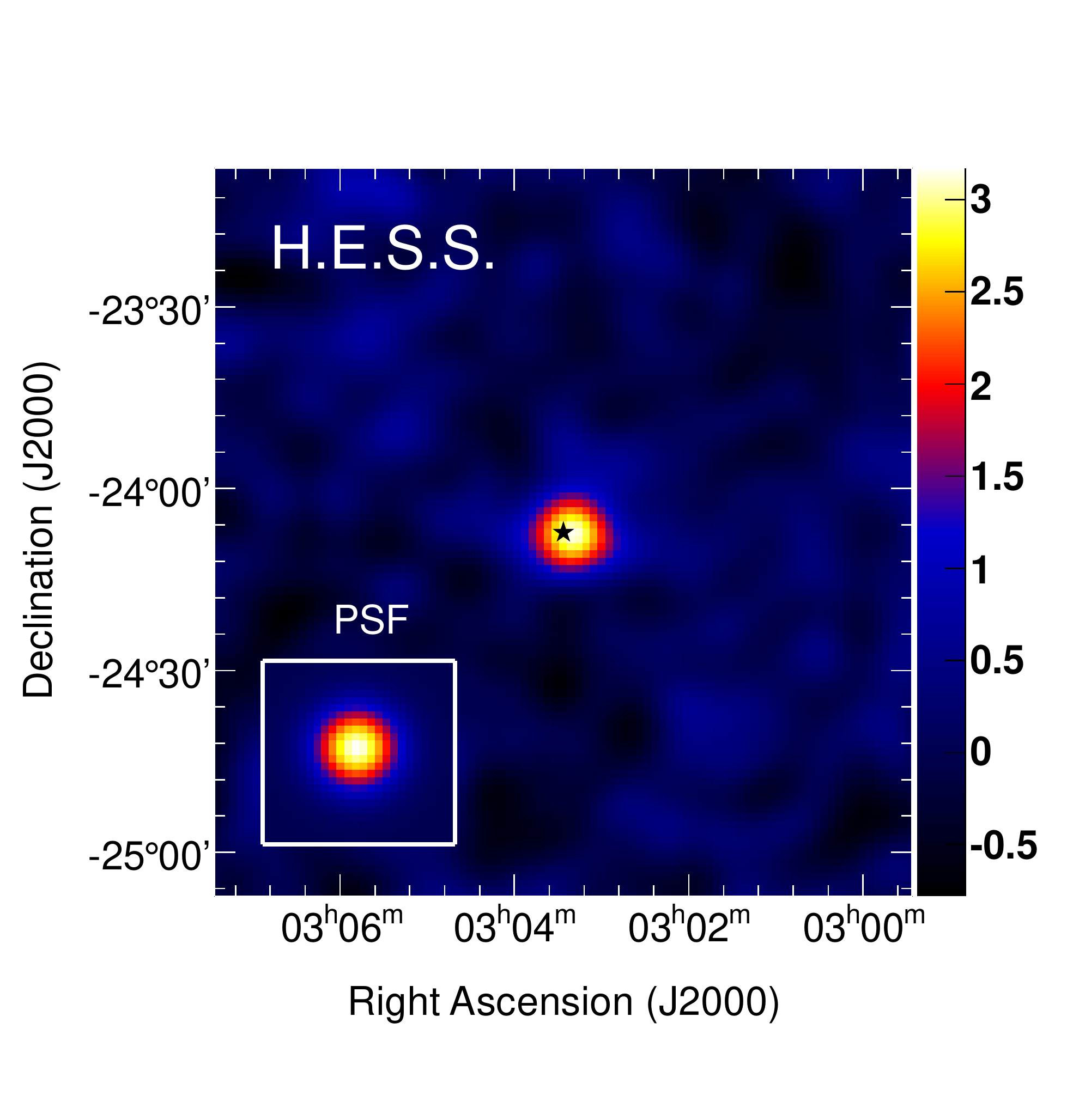}
\caption{Smoothed \g-ray excess map of H.E.S.S. events in units of counts per arcmin$^2$ centred on the position of \pks (see text for details). The PSF is shown in the inset. The star marks the position of \pks as measured in the infrared. }
\label{fig:map}
\end{figure}

The analysis is performed with \textit{Standard cuts} for an efficient background rejection \citep{2009APh....32..231D}. An excess of 264 \g-ray candidates (900 ON events, 7638 OFF events, background normalization 0.083) in a circular region of radius $0.1\degr$ centred on \pks is measured. This excess corresponds to a significance of 9.4$\sigma$, using Eq.~17 from \citet{1983ApJ...272..317L}. The smoothed excess map is shown in Fig.~\ref{fig:map}. The map is smoothed with a Gaussian with a width of $3.5\arcmin$. This width corresponds to the 68\% confinement radius of the point spread function (PSF) for this analysis. The excess is found to be point-like within the statistical uncertainties. A fit of the uncorrelated excess map with the PSF of the instrument gives a position for the excess of $\alpha_\mathrm{J2000}=03^\mathrm{h} 03^\mathrm{m} 23\fs49 \pm 1\fs19_\mathrm{stat} \pm 1\fs30_\mathrm{syst}$, $\delta_\mathrm{J2000}=-24\degr 07\arcmin 35\farcs86 \pm 15\farcs35_\mathrm{stat} \pm 19\farcs50_\mathrm{syst}$, consistent at the 1$\sigma$ level with the nominal position of \pks ($\alpha_\mathrm{J2000}=03^\mathrm{h} 03^\mathrm{m} 26\fs49$, $\delta_\mathrm{J2000}=-24\degr 07\arcmin 11\farcs50$) reported by \cite{2003yCat.2246....0C}.
 
\begin{figure} 
  \centering
  \includegraphics[width=0.8\columnwidth]{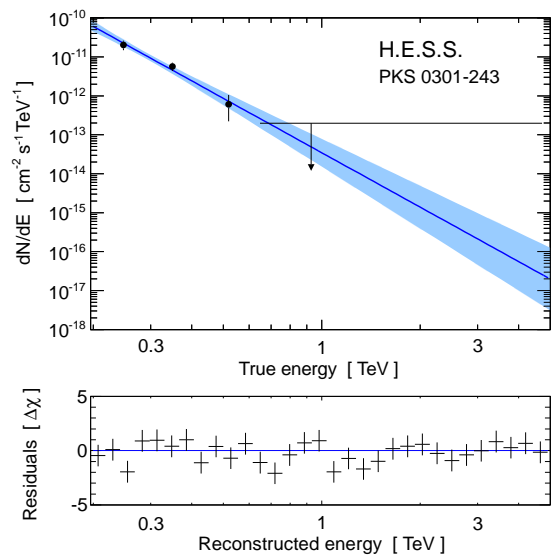}
  \caption{Forward folded spectrum of \pks. Top panel: the blue line is the best fit of a power law to the data as a function of the true energy (unfolded from \hess response functions). The points are an unfolded representation of the data assuming the best fit spectrum. The blue bow tie plot is the uncertainty of the fit given at a confidence level (C.L.) of 1$\sigma$. Upper limits are given at 3$\sigma$ C.L. with \cite{1998PhRvD..57.3873F} confidence intervals. Bottom panel: residuals of the fit normalized to the errors as a function of the reconstructed (measured) energy. The blue line corresponds to the no-deviation case.}
  \label{fig:spectrum}
\end{figure}

Figure~\ref{fig:spectrum} shows the differential VHE \g-ray spectrum of \pks above the energy threshold of 200 GeV, computed using a forward folding technique \citep{2001A+A...374..895P}. In this technique, a likelihood estimator is built assuming that the number of counts in each reconstructed energy bin is Poissonian. The reconstructed energy bins show the energy of the events as it is measured by \hess The most likely values of the parameters for the assumed spectral shape are retrieved by comparing the observed number of counts in the reconstructed energy bins to the number of counts in the same bins expected from the theoretical spectrum. The spectrum is well described by a power-law shape ($dN/dE \propto (E/E_{\rm d})^{-\Gamma}$). The equivalent $\chi^2$ per number of degree of freedom  $n_\mathrm{dof}$ is $\chi^2/n_\mathrm{dof} = 35.2/29$. The photon index is $\Gamma = 4.6 \pm 0.7_{\mathrm{stat}} \pm 0.2_{\mathrm{syst}}$ and the decorrelation energy $E_{\rm d}$ is 290 GeV. The integral flux above 200 GeV is $I(E>200\,\mathrm{GeV}) = (3.3 \pm 1.1_\mathrm{stat} \pm 0.7_\mathrm{syst}) \times 10^{-12}$\,\phcms. This flux corresponds to 1.4\%  of the Crab Nebula flux above the same energy threshold \citep{2006A+A...457..899A}. A log-parabola ($dN/dE \propto (E/E_{\rm d})^{-\alpha-\beta\log(E/E_{\rm d})}$) or a power law with an exponential cut-off  ($dN/dE \propto (E/E_{\rm d})^{-\Gamma}e^{-E/E_{\rm c}}$) do not significantly improve the fit of the spectrum.

The light curve of the integrated flux for $E>200$\,GeV for the different periods of observation is shown in Fig.~\ref{fig:lc}. The fit of a constant to the data yields  a $\chi^2$ of 19.2 for 8 degrees of freedom, corresponding to a probability of 0.014. This probability corresponds to a hint for variability at the $2.5\sigma$ level. The amplitude of intrinsic variation of the flux can be estimated as in \citet{2003MNRAS.345.1271V} by calculating the fractional variance. The fractional variance is defined as the square root of the excess variance, thus accounting for the intrinsic scatter of fluxes that is not due to shot noise. For the VHE light curve of \pks,  $F_\mathrm{var} = (23\pm27) \%$ is found, where $27\%$ is the amplitude of variation induced by random Poisson processes.

\section{Multi-wavelength observations}
\label{sec-mwl}

\subsection{Fermi/LAT observations}
\label{subsec-fermi}

The LAT \citep{2009ApJ...697.1071A} is a pair conversion telescope onboard the \textit{Fermi} satellite that was launched in June 2008. It is sensitive to \g-rays between 20 MeV and a few hundred GeV. The primary mission of the instrument is to make a \g-ray survey, the full sky being covered every three hours.

Previous analyses of LAT data of \pks showed evidence for a flaring episode of the source in April/May 2010 \citep{2010ATel.2591....1C,2010ATel.2610....1N} and a detection at VHE using LAT data above 100 GeV \citep{2011A+A...529A..59N}. Moreover, \pks was associated in the {\it Fermi} Bright Source List \citep{2009ApJS..183...46A} with \object{0FGL\,J0303.7$-$2410}, as well as with \object{1FGL\,J0303.5$-$2406} in the first source catalogue \citep{2010ApJS..188..405A}, and with \object{2FGL\,J0303.4$-$2407} in the second source catalogue \citep[2FGL,][]{2012ApJS..199...31N}. In the 2FGL, \pks is reported as being detected at 47.0$\sigma$ with an energy flux of $F=(7.66 \pm 0.42) \times 10^{-11}$\,\ergcms and a photon index of $\Gamma=1.94 \pm 0.03$ in the 100\,MeV--100\,GeV energy range.

\begin{table*}
  \caption{Spectral parameters from the \fermi likelihood analysis. LLRT is the log-likelihood ratio test.}
  \label{tab:specparam}  
  \centering
  \begin{tabular}{cccccccc}
    \hline\hline
           & Hypothesis & TS      & First parameter & Second parameter  &  $F_{0.2-300\,\mathrm{GeV}} (10^{-8}\,$\phcms$)$ & LLRT / Prob.\\
    \hline
           & PL         & 2236.13 & 1.94 $\pm$ 0.03 & --                & 2.08 $\pm$ 0.10 & -- \\
Low state  & BPL        & 2239.15 & 1.69 $\pm$ 0.18 & 1.98  $\pm$ 0.04  & 1.97 $\pm$ 0.13 & 2.15 / 22.6\% \\
           & LP         & 2236.91 & 1.92 $\pm$ 0.04 & 0.013 $\pm$ 0.017 & 2.04 $\pm$ 0.11 & 0.53 / 39.2\% \\
    \hline
           & PL         & 1190.20 & 1.86 $\pm$ 0.05 & --                & 34.3 $\pm$ 2.5  & --\\
High state & BPL        & 1193.52 & 1.41 $\pm$ 0.39 & 1.91  $\pm$ 0.07  & 33.4 $\pm$ 5.2  & 1.54 / 26.6\%\\
           & LP         & 1193.01 & 1.80 $\pm$ 0.07 & 0.041 $\pm$ 0.034 & 33.6 $\pm$ 2.5  & 1.56 / 26.4\%\\
    \hline
  \end{tabular}
  \tablefoot{The first parameter corresponds to the photon index $\Gamma$ for a power-law (PL) hypothesis or $\alpha$ for a log-parabola (LP) hypothesis, or to the first photon index $\Gamma_1$ for a broken power-law (BPL) hypothesis. The second parameter is either the second photon index $\Gamma_2$ in the case of a BPL hypothesis or the curvature parameter $\beta$ for a LP hypothesis. For the LP model $E_\mathrm{b}$ is fixed at 1 GeV. The break energy for the BPL is, respectively, 0.60 $\pm$ 0.15 GeV and 0.47 $\pm$ 0.33 GeV in the low and high state.}
\end{table*}

Public LAT data\footnote{\url{http://fermi.gsfc.nasa.gov/ssc/data/access}} from August 4, 2008 (MJD\,54682) to October 1, 2012 (MJD\,56201) are analysed here using the \textit{Fermi Science Tools}\footnote{\url{http://fermi.gsfc.nasa.gov/ssc/data/analysis/software}} \texttt{v9r27p1}, and the \texttt{P7SOURCE\_V6} instrumental response functions. Light curves and spectra are produced using a binned likelihood analysis by selecting \texttt{SOURCE} class events with energies between 200 MeV and 300 GeV, in a circular region of interest (RoI) of 10\degr\ of radius around the nominal position of \pks. Cuts are applied on the zenith angle with respect to the Earth ($<100\degr$), and on the rocking angle of the spacecraft ($<52\degr$). All the objects included in the 2FGL \citep{2012ApJS..199...31N} within 15\degr\ of the RoI centre are included in the model construction of \pks. The isotropic model \texttt{iso\_p7v6source} is used to account for both the extragalactic diffuse emission and the residual instrumental background, while the spatial template \texttt{gal\_2yearp7v6\_v0} is used to account for the contribution from the Galactic diffuse emission.

A high state data set has been defined, lasting from MJD\,55312 (April 26, 2010) to MJD\,55323 (May 5, 2010), corresponding to the peak of the flare. Low state events are retained from MJD\,54683 (August 5, 2008) to MJD\,55251 (February 24, 2010) and from MJD\,55351 (June 4, 2010) to MJD\,56201 (October 1, 2012). The light curve analysis presented below shows that the flux after the flare is already low at MJD\,55351 so that this date can safely be considered for the beginning of the second time window used to define the low state. In the high state, \pks is detected with a test statistic \citep[TS,][]{1996ApJ...461..396M} of 1190.20, approximately corresponding to 34$\sigma$, while it is detected with TS $=2236.13$ ($\approx 47\sigma$) in the low state of activity. The spectra for both high and low states are well described by a power-law shape. Spectral parameters for the low state and the high state of activity are summarised in Table ~\ref{tab:specparam}. The corresponding energy flux in the 200\,MeV--300\,GeV energy range for the low state is $F_{0.2-300\,\mathrm{GeV}}=(5.71 \pm 0.40) \times 10^{-11}$\,\ergcms with $\Gamma=1.94 \pm 0.03$, and $F_{0.2-300\,\mathrm{GeV}}=(1.22 \pm 0.19) \times 10^{-9}$\,\ergcms with $\Gamma=1.86 \pm 0.05$ for the flaring state. Table \ref{tab:specparam} shows that the photon index is slightly different between the flaring state and the quiescent state. However, no significant correlation over all the bins of the light curve between the photon index and the integral flux is observed. In this table, log-likelihood ratio test (LLRT) values are also reported, comparing a broken power law
\begin{equation*}
  \frac{dN}{dE} = N_0 \times\left\{
  \begin{array}{ll}
    (E/E_\mathrm{b})^{\Gamma_1} & \mbox{if $E < E_\mathrm{b}$}\\
    (E/E_\mathrm{b})^{\Gamma_2} & \mbox{otherwise}
  \end{array} \quad\mbox{\cmsmev} \right.
\end{equation*}
or a log-parabola spectral hypothesis with respect to the simple power-law hypothesis. The equivalent $\chi^2$ probabilities for the null hypothesis\footnote{i.e. that the power-law hypothesis describes the spectrum of \pks better.}, following Wilks' theorem \citep{1938AnMathStat.9.60W}, are also reported, and show that neither the log-parabola nor the broken power-law hypothesis gives a significantly improved fit to the data, compared to a power-law spectrum, both in quiescent and flaring state of \pks in the HE range. 

The light curve computed using 10-days integration bins is displayed in Fig.~\ref{fig:lc}. An integration time of ten days for the time binning is chosen to ensure minimum statistics in each bin. The light curve shows a pronounced flaring episode between MJD\,55306 (April 20, 2010) and MJD\,55329 (May 13, 2010). At the maximum of the flare, around MJD\,55314 (April 28, 2010), $F_\mathrm{0.2-300 \,GeV}=(7.01 \pm 1.16) \times 10^{-7}$\,\phcms is 34 times the base flux in the low state, $F_\mathrm{0.2-300 \,GeV}=(2.08 \pm 0.10) \times 10^{-8}$\,\phcms. The fit of the data points in the low state period with a constant ($\chi^2/n_{dof} = 215/89$), not taking into account upper limits, shows evidence for variability in the low state.  The measured fractional variability constrains the amplitude of intrinsic variability to $F_\mathrm{var} = (14.5 \pm 11.2)\%$ where $11.2\%$ is the fractional variability induced by random Poisson processes. A minimal variability timescale can be estimated by using the method of doubling-time \citep[see e.g. ][]{1992ApJ...401..516E, 1999ApJ...527..719Z}. As in \cite{1999ApJ...527..719Z}, a linear interpolation between each group of two points is used to estimate the time corresponding to a doubling of the flux. This quantity depends, however, on the sampling of the light curve and on the signal to noise ratio. 
Conservatively, the shortest timescale of variability is defined as the average over the five lowest values having an uncertainty less than 30\%~\citep{2000ApJ...541..153F}. Here, variability down to a timescale of ten days is found, which corresponds to the integration time used in the light curve. While the source may exhibit faster variability, this cannot be probed in the present data set, since the use of smaller time bins in the time series analysis would increase the uncertainties of the measurements. If present in the VHE light curve, this small amplitude of variation cannot be detected by \hess given the weakness of the source and the sparse sampling of the VHE light curve.

A dedicated analysis of the whole period in the 100\,MeV--100\,GeV band, for a better comparison with results from the 2FGL catalogue, yields an energy flux of $F_{0.1-100\,\mathrm{GeV}} = (6.41 \pm 0.28) \times 10^{-11}$\,\ergcms with $\Gamma=1.93 \pm 0.02$. These results, compared to those reported in the 2FGL catalogue, show a fully compatible photon index but a slightly lower flux. This slight difference in the flux can be understood by the fact that more data were used in the analysis presented here.
Since the long-term light curve reveals only one important active event, in April-May 2010, the average flux is lowered by the two additional years of low-state integration in 2011 and 2012.

\subsection{Swift/XRT and UVOT observations}
\label{subsec-swift}

The object \pks was observed with \swift \citep{2004ApJ...611.1005G} between 2009 and 2012 in seven pointed observations, for a total exposure of 25.1\,ks. These data were analysed using the package \texttt{HEASOFT 6.12}.

The XRT is a focusing X-ray telescope with grazing incidence mirrors using a CCD imaging spectrometer for detection. It is sensitive to X-rays between 0.2 and 10 keV with a PSF of 18$\arcsec$ at 1.5 keV.  Only data taken in photon counting mode are considered here. Data are recalibrated using the standard procedure \texttt{xrtpipeline}. Source events are selected within a circle with a radius of 20 pixels ($0.79\arcmin$) centred on the nominal position of \pks and background events are extracted from an annular region of 50 pixels ($1.97\arcmin$) to 120 pixels ($4.72\arcmin$) centred on the source. The observations are individually checked for pile-up effects, which is found to be negligible.

\begin{figure*}
  \centering
  \includegraphics[width=\textwidth]{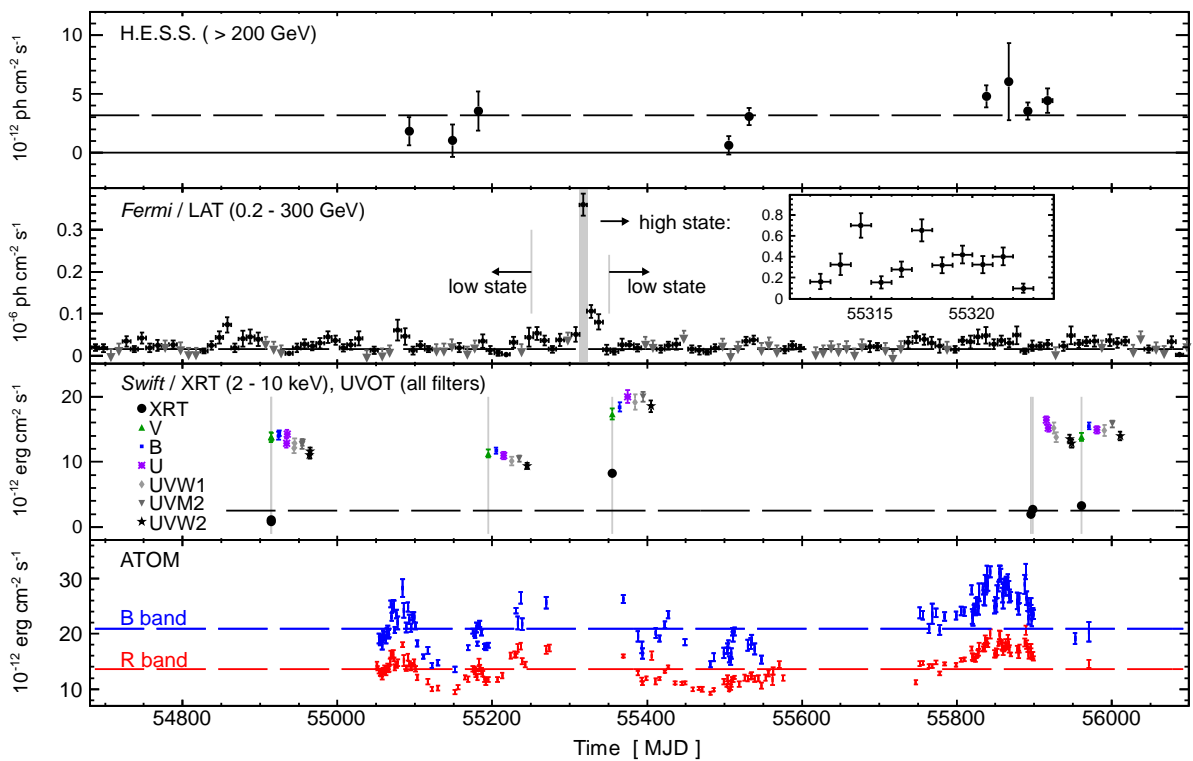}
  \caption{Multi-wavelength light curves for \pks for the period MJD\,54600 (May 14, 2008) to MJD\,56201 (October 1, 2012). From top to bottom: Panel 1: VHE light curve from \hess data, the dashed line is the average integrated flux over all observations. The solid line corresponds to a null flux. Panel 2: LAT data with 10 days integration bins, the inset is a zoom on the high-state period with one-day integration bins. The dashed line is the best fit of the low state points to a constant. Panel 3: \swift/XRT observation light curve for integrated flux between 2 and 10 keV. The error bars for the \swift/XRT measurements are too small to be distinguishable in this plot. No data are shown around MJD 55195 (December 30, 2009) because of insufficient statistics for a spectral fit. The dashed line is the best fit of these data to a constant. The light curves of UVOT data for all filters are also shown. For clarity, for each pointed observation (marked with grey solid lines), the UVOT results for the different filters are shifted along the increasing time axis, following the order V, B, U, UVW1, UVM2, UVW2.  Panel 4: ATOM light curve in B and R bands, with one exposure per night contemporaneous with \hess observations. The dashed horizontal lines are the best fit of a constant to the data.}
  \label{fig:lc}
\end{figure*}

The spectral analysis above 0.3 keV is performed using the package \texttt{XSPEC 12.7.1}. Spectra are binned to ensure a minimum of 30 counts per bin so that the number of counts in each bin follows a Gaussian distribution. The Galactic absorption is accounted for with a hydrogen column density fixed to $N_\mathrm{H} = 1.70 \times 10^{20}$\,cm$^{-2}$ \citep{2005A+A...440..775K}. Data from each observation are well fitted by an absorbed power-law spectrum. Broken power-law and log-parabola spectral shapes do not improve the fit of the \swift/XRT data, for any exposure. Spectral parameters for each fit are given in Table~\ref{tab-swift}. The spectrum for the total exposure is also well fitted by an absorbed power law spectrum ($\chi^2/n_\mathrm{dof} = 81.8/82$), with parameters shown in the same table. A fit by an absorbed power-law with free hydrogen column density on the sum of all observations yields a value of $N_\mathrm{H}=(1.64 \pm 0.7) \times 10^{20}$\,cm$^{-2}$, compatible with the fixed value used throughout the analysis.

The light curve of the integral flux between 2 and 10 keV, represented in Fig.~\ref{fig:lc}, shows significant variability, with a $\chi^2/n_{\rm dof}$ = 330/5 for a fit with a constant flux and a fractional variance $F_{\rm var} = (82 \pm 3)\%$. No significant variation of the spectral index between the various observations is found.

Contemporaneously with XRT data, UVOT observations were made using the six filter settings available. The UVOT instrument \citep{2005SSRv..120...95R} onboard \swift measures the UV emission in the bands V (544 nm), B (439 nm), U (345 nm), UVW1 (251 nm), UVM2 (217 nm), and UVW2 (188 nm) simultaneously with the X-ray emission. These data have been recalibrated and the instrumental magnitudes and the corresponding fluxes (see \citealt{2008MNRAS.383..627P} for details on the calibration procedure) are calculated with {\tt uvotsource} taking into account all photons from a circular region with a radius of $5\arcsec$ (standard aperture for all filters). It is assumed that the count rate to flux conversion factors, computed for a mean GRB spectrum, are applicable in the case of PKS 0301-243. An appropriate background is determined from a circular region near the source region without contamination from other sources. The magnitudes measured at each epoch and for all filters used are shown in Table~\ref{tab-uvot}. The light curves for all filters is displayed in Fig.~\ref{fig:lc}. Just as the X-ray band light curve did, these curves show pronounced variability with, however, a smaller amplitude of variation. In the U band, a fit with a constant has a $\chi^2/n_{\rm dof}$ = 89/6 and a fractional variance  $F_{\rm var} = (17 \pm 1) \%$ is measured. X-ray and optical bands show a hint of correlation with a linear Pearson correlation coefficient of $r = 0.85 \pm 0.14$ between the X-ray and the U band, for 4 degrees of freedom. This corresponds to a probability of non-correlation of $0.03$.

\begin{table*}
\caption{Parameters deduced from the spectral analysis of \swift/XRT data.}
\label{tab-swift}
\centering
\begin{tabular}{cccccc}
\hline\hline
Observation date & Observation ID & Exposure time (s) & $F_{2-10\,\mathrm{keV}}$\tablefootmark{a} ($10^{-12}\,$\ergcms) & $\Gamma$        & $\chi^2$/dof\\
\hline
24/03/2009       & 00038098001    & 1883              & $0.81^{+0.28}_{-0.26}$                      & $2.68 \pm 0.32$ & 1.00/3\\[3pt]
24/03/2009       & 00038368001    & 4912              & $1.09^{+0.17}_{-0.12}$                      & $2.52 \pm 0.14$ & 9.72/15\\[3pt]
30/12/2009\tablefootmark{b} & 00038098002 & 912       & --                                       & --              & --\\[3pt]
07/06/2010       & 00038098003    & 4715              & $8.27^{+0.42}_{-0.43}$                       & $2.30 \pm 0.05$ & 113.4/72\\[3pt]
30/11/2011       & 00038098004    & 3751              & $1.98^{+0.19}_{-0.17}$                       & $2.57 \pm 0.11$ & 21.7/23\\[3pt]
03/12/2011       & 00038098005    & 3938              & $2.67^{+0.20}_{-0.28}$                       & $2.49 \pm 0.09$ & 17.5/25\\[3pt]
04/02/2012       & 00038098006    & 5022              & $3.30^{+0.15}_{-0.23}$                       & $2.40 \pm 0.07$ & 40.0/41\\[3pt]
\multicolumn{2}{c}{All observations} & 25133         & $3.11^{+0.16}_{-0.13}$                       & $2.44 \pm 0.03$ & 81.8/82\\[3pt]
\hline
\end{tabular}
\tablefoot{
\tablefoottext{a}{not corrected for Galactic absorption.}\\
\tablefoottext{b}{not enough data to allow a spectral fit.}
}
\end{table*}

\begin{table*}
\caption{Magnitudes\tablefootmark{a} at different epochs from \swift/UVOT data in all bands.}
\label{tab-uvot}
\centering
\begin{tabular}{ccccccc}
\hline\hline
Observation ID & V & B & U & UVW1 & UVM2 & UVW2\\
\hline
00038098001    & $ 15.43 \pm 0.07 $  & $ 15.77 \pm 0.06 $  & $ 14.95 \pm 0.06 $ & $ 14.83 \pm 0.07 $ & $ 14.79 \pm 0.07 $  & $ 14.92 \pm 0.07 $\\[3pt]
00038368001    & $ 15.43 \pm 0.05 $  & $ 15.74 \pm 0.05 $  & $ 14.83 \pm 0.05 $ & $ 14.76 \pm 0.06 $ & $ 14.76 \pm 0.06 $  & $ 14.87 \pm 0.06 $\\[3pt]
00038098002    & $ 15.65 \pm 0.06 $  & $ 15.96 \pm 0.06 $  & $ 15.12 \pm 0.06 $ & $ 15.03 \pm 0.07 $ & $ 14.99 \pm 0.07 $  & $ 15.10 \pm 0.06 $\\[3pt]
00038098003    & $ 15.18 \pm 0.06 $  & $ 15.47 \pm 0.06 $  & $ 14.47 \pm 0.06 $ & $ 14.33 \pm 0.06 $ & $ 14.29 \pm 0.06 $  & $ 14.36 \pm 0.06 $\\[3pt]
00038098004    & --\tablefootmark{b} & --\tablefootmark{b} & $ 14.68 \pm 0.05 $ & $ 14.58 \pm 0.06 $ & --\tablefootmark{b} & $ 14.71 \pm 0.06 $\\[3pt]
00038098005    & --\tablefootmark{b} & --\tablefootmark{b} & $ 14.76 \pm 0.05 $ & $ 14.69 \pm 0.06 $ & --\tablefootmark{b} & $ 14.76 \pm 0.06 $\\[3pt]
00038098006    & $ 15.43 \pm 0.05 $  & $ 15.66 \pm 0.05 $  & $ 14.83 \pm 0.05 $ & $ 14.61 \pm 0.06 $ & $ 14.55 \pm 0.06 $  & $ 14.67 \pm 0.06 $\\[3pt]
\hline
\end{tabular}
\tablefoot{
\tablefoottext{a}{not corrected for Galactic extinction.}\\
\tablefoottext{b}{no observations were taken in these filters for these observation IDs. For the other reported measurements, the exposure varies for the different filters and observations but is of the order of a few hundred seconds in these cases.}\\
}
\end{table*}

\begin{figure}
  \centering
  \includegraphics[width=0.8\columnwidth]{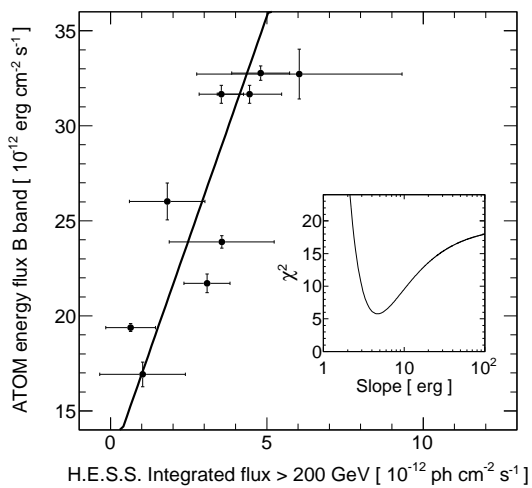}
  \caption{Nine simultaneous measurements of energy flux in R band by ATOM and integrated flux above 200 GeV by \hess The line is the best fit of a linear function to the data. The inset shows the $\chi^2$ profile when varying the slope of the linear law.}
  \label{fig:atom_hess}
\end{figure}

\subsection{ATOM observations}
\label{subsec-atom}

The Automatic Telescope for Optical Monitoring (ATOM) is a 75 cm optical telescope operated at the \hess site in Namibia \citep{2004AN....325..659H} and has been monitoring \pks since 2009, mainly in the R and B bands with 600s exposures in the R band and 800s or 900s in the B band. Observations are taken at a cadence of one frame per night contemporaneous with \hess observations. Fluxes are calculated using a $4\arcsec$ radius aperture, and the light curve in the R and B bands is shown in Fig.~\ref{fig:lc}. The flux is highly variable on a timescale of a dozen days with a variability of $F_{\rm var} = (18.6 \pm 0.4)\%$ in the B band and $F_{\rm var} = (17.9 \pm 0.3)\%$ in the R band. Given the good time-sampling of the light curve, a minimal variability timescale can be computed using the same doubling-time method used for the HE light curve.  In both B and R bands, the shortest variability timescale found is 8 days, of the same order as the variability timescale found in the LAT light curve. To reduce the bias on the variability timescale arising from the uneven sampling of the ATOM light curve, the shortest variability timescale is conservatively defined as in the LAT analysis.

One of the missions of ATOM is to monitor sources simultaneously with \hess Here, ATOM exposures have been taken during all the periods of \hess observations of \pks. The correlation between simultaneous ATOM and \hess observations is probed by averaging all ATOM exposures that were made in the corresponding \hess observation periods. The linear Pearson correlation coefficient found is 0.84 $\pm$ 0.18 between \hess and B-band data and 0.85 $\pm$ 0.17 between \hess and R-band data. This corresponds to a probability of non-correlation of 0.005 and 0.004 respectively, indicating that the optical and the VHE bands may be correlated. Figure \ref{fig:atom_hess} shows the averaged energy flux in the B band measured by ATOM during the nine \hess observation periods as a function of the corresponding integrated flux above 200 GeV. These data are well fitted by a linear law ($\chi^2/n_{\rm d.o.f.} = 5.7/7$) with a finite slope of 4.7 erg. The inset in Fig.~\ref{fig:atom_hess} shows the $\chi^2$ profile when varying the slope of the linear law.  The $\chi^2$ difference between the best-fit value and the $\chi^2$ for an infinite slope shows that a finite value for the slope is preferred at the 3.5$\sigma$ level, confirming that the two bands may be correlated. Strong correlations between these two bands have been observed for the first time for an HBL in a low state of activity of \object{PKS\,2155$-$304} \citep{2009ApJ...696L.150A}. The correlation between optical and HE bands using LAT data is hard to probe given the long integration time required to form time bins in the HE light curve that are significant. However, the minimal timescale of variability of 8 days found in the  ATOM data is of the same order as the integration time used to produce the HE light curve. For this reason, the correlation between ATOM and \fermi light curves has been estimated by averaging ATOM measurements in the same time binning as the HE light curve. This way, 31 simultaneous bins are formed. No significant correlation is found, with a coefficient $r = 0.14 \pm 0.17$ corresponding to a probability of non-correlation between the optical and the HE light curve of 0.45.

\begin{figure*}
  \centering
  \includegraphics[width=0.8\textwidth]{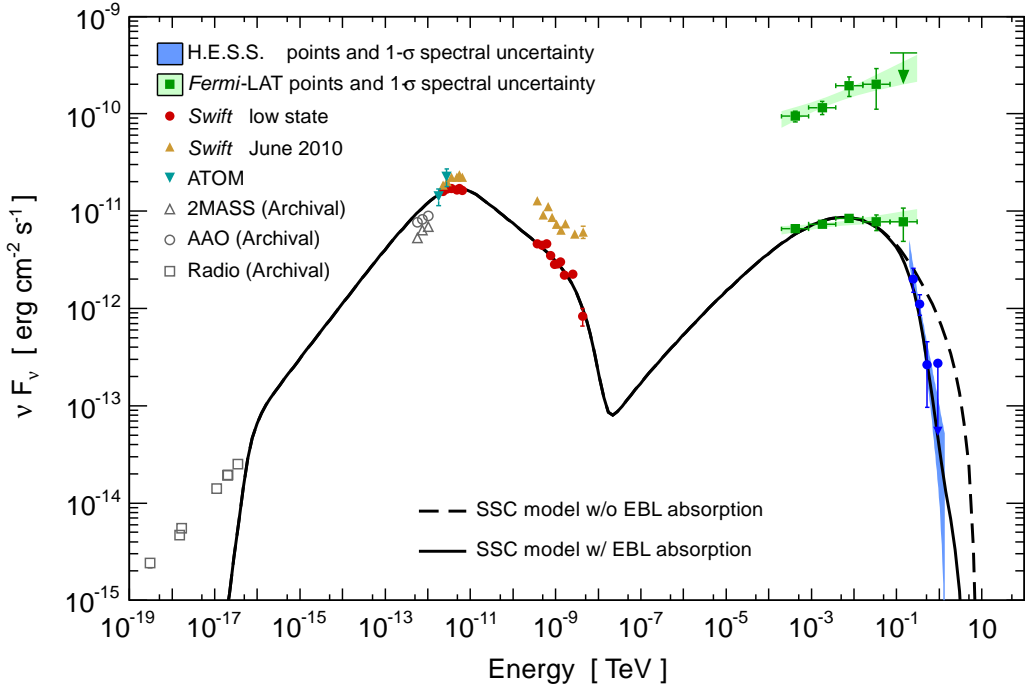}
  \caption{Spectral energy distribution of \pks. Shown are the \hess spectrum with uncertainty given at 1$\sigma$ C.L., bow tie plots corresponding to the \fermi $1\sigma$  spectral uncertainties for both low state and flaring state with measurements, ATOM data in R and B bands where the error bars represent the measured variability (see text for details), and \swift/UVOT and XRT data for the low state and the 2010 observations. Archival data in radio and infrared extracted from NED are also shown. Data in X-ray and optical bands are corrected for Galactic absorption. The black lines correspond to the SSC model with (thick) and without (dashed) EBL absorption. The EBL model is taken from \citet{2008A+A...487..837F}.}
  \label{fig:sed}
\end{figure*}

\section{Discussion}
\label{sec-discussion}

\subsection{Interpretation of the multi-wavelength spectral energy distribution}
\label{subsec-mwlsed}

The overall spectral energy distribution (SED) is shown in Fig.~\ref{fig:sed}, including the \hess spectrum, and bow tie plots for LAT $1\sigma$ spectral uncertainties in both low and flaring states. Averaged fluxes from ATOM data in R and B bands are also displayed. These points are the average over all the measurements shown in Fig. \ref{fig:lc} for the two bands. The error bars are the r.m.s. of the measurements, thus accounting for the measured flux variability. Fluxes are corrected for Galactic absorption with a reddening of 0.022 mag \citep{1998ApJ...500..525S}. Also shown are infrared data (Two Micron All-Sky Survey, United Kingdom Infra-Red Telescope, and the Australian Astronomical Observatory) respectively from \cite{2005NewA...11...27C,1982MNRAS.199..969A,1983MNRAS.205..793W} and radio data extracted from the NASA extragalactic database archive \citep{2007AJ....134.1245C,1996AJ....111.1945D,1994ApJS...90..179G,2010MNRAS.402.2403M,1990PKS...C......0W}. Radio data are measurements of the total integrated fluxes.  The points in infrared, optical, and X-ray bands have been corrected for Galactic absorption. The emission by the host galaxy in optical is expected to contribute to approximately 4\% of the average optical flux measured by ATOM in the R band, or even less in the \swift/UVOT range, using a giant elliptical galaxy template \citep{2001MNRAS.326..745M}. \cite{2000A+A...357...91F} found the radial profile of the source in the R band to be well modelled by a point source plus a faint elliptical component contributing to 2\% of the total flux. This contribution is therefore neglected in the following discussion.

{\it Swift} data in X-ray and optical/UV have been divided into two states of activity of the source. The high state, following the flare in the LAT energy range, includes only the observations in June 2010 for which the measured X-ray flux is four times the time-averaged flux of the remaining observations. For the low state, non-simultaneity between the \hess, LAT and {\it Swift} observations could weaken the interpretation of a multi-wavelength SED. Nevertheless, the limited amplitude of the broad-band variability  suggests that the low-state spectra can be interpreted together despite non-simultaneity. This state includes the \hess observations, LAT data before and after the 2010 flare (see Sect.~\ref{subsec-fermi}) and \swift data excluding the June 2010 observations.

The SED of \pks presents two broad peaks, which is a general feature of BL\,Lac objects. For \pks, the low energy component peaks in optical or near ultraviolet wavelengths. From the two \swift spectra in Fig.~\ref{fig:sed}, the location of the peak, on average at $\nu_s \sim 10^{15} \rm Hz$ at the formal boundary between IBL and HBL, seems to vary with the level of flux, showing a tendency of higher frequencies during higher fluxes. This correlation has already been observed in blazars \citep[e.g.][]{2000ApJ...541..166F,2004ApJ...601..759T} and is a prediction of some particle acceleration models in jets of VHE blazars \citep[see e.g.][]{2006A+A...453...47K}. To probe this behaviour, the location of the peak for the two states has been estimated using UVOT data. A log-parabola is fitted to the de-reddened fluxes to deduce the position and flux of the peak. For the low state of activity, which is an average of different observations, the low energy peak  is at a frequency $\log_{10}(\nu_\mathrm{s}/1 \rm Hz)= 14.95 \pm 0.03$, with an energy flux $F_\mathrm{s} = (1.56 \pm 0.30) \times 10^{-11} $ \ergcms. For the high state, a peak frequency of $\log_{10}(\nu_\mathrm{s}/1 \rm Hz) = 15.19 \pm 0.05$ is found, with an energy flux of $(2.34 \pm 0.50) \times 10^{-11} $ \ergcms. Between these two states, the peak is therefore significantly shifted to higher energies for higher fluxes.

The simplest and most common model used to account for the two peaks of the SEDs of BL Lac objects is the synchrotron self-Compton (SSC) model \citep[see e.g.][]{1965ARA+A...3..297G}. The SSC model used here \citep{2001A+A...367..809K} describes a spherical emitting region of electrons filled with a magnetic field $B$ and propagating relativistically with a Doppler factor $\delta$. Here, the energy spectrum of the electrons is assumed to follow a broken power-law shape with a break energy $\gamma_\mathrm{b}$. In the framework of this model, the SED component at lower energies comes from synchrotron radiation of electrons in the magnetic field of the emitting region. The other component at higher energies results from inverse-Compton scattering of the synchrotron photons on the high energy electrons. The parameters of the SSC model can be constrained by observations \citep{1996A+AS..120C.503G, 1998ApJ...509..608T}. The location of the synchrotron and inverse-Compton (IC) peaks, that are here well sampled by multi-wavelength data, can be used to put a constraint on $B\delta$ through the relation \citep{1998ApJ...509..608T}
\begin{equation*}
 B\delta = (1+z) \frac{(\nu_\mathrm{s}/\mathrm{1 Hz})^2}{2.8\cdot 10^6 (\nu_{\rm IC}/\mathrm{1 Hz})} \; \mathrm{[G]}\;\; , 
\end{equation*}
where $\nu_s$ is the frequency of the synchrotron peak, taken here from the low state data, $\nu_{\rm IC}$ is the frequency of the IC peak, and $z = 0.266$ is the redshift of the source. This relation holds as long as the IC emission at the peak lies in the Thomson regime. This asymptotic regime is verified for energies of the back-scattered photon smaller than $m_\mathrm{e}^2c^4\delta^2/4h\nu_\mathrm{s}(1+z)^2$~\citep{1998ApJ...509..608T}. To ensure that the IC emission lies in the Thomson regime up to 5 GeV, corresponding to the energy of the IC peak, $\delta$ should be larger than 1.3.

The position of the synchrotron peak has been determined above. In a similar fashion, the position of the IC peak can be estimated by fitting a log-parabola on the HE points from \hess and \fermi. Absorption on the EBL is taken into account for the \hess points using the model of \cite{2008A+A...487..837F}. This results in a position of the IC peak of $\log_{10}(\nu_{\rm IC}/ 1\rm Hz) = 23.93 \pm 0.15$ for an energy flux $F_{\rm IC} = (7.91 \pm 0.40) \times 10^{-12} $ \ergcms. This position is fully compatible with the empirical relation between the IC peak position and the photon index measured by \fermi that was deduced on the basis of a sample of 48 blazars, not including \pks \citep[see Eq. (5) of][]{2010ApJ...716...30A}: the peak position from that relation, taking a photon index of 1.94, is $\log_{10}(\nu_{\rm IC}/ 1\rm Hz) = 23.84 \pm 0.70$. The value of $B\delta$ for the SSC model should be in the range $\log_{10}(B\delta/1\rm G) = -0.37 \pm 0.16$.

A second and independent constraint on $B$ and $\delta$ can be constructed from the minimal variability timescale $t_{\rm var}$ that is measured. In the following, the value of 8 days found in the ATOM light curve will be used as an upper limit on $t_{\rm var}$. For causality reasons, the emitting region radius $R_b$ cannot be larger than $ct_{\rm var}\delta/(1+z)$. This condition translates into a lower limit on $B\delta^3$ as \citep{2008ApJ...686..181F}
\begin{equation*}
B\delta^3 \geq \sqrt{\frac{24\pi}{c^3}}\frac{d_{\rm L}(1+z)}{t_{\rm var}}\frac{F_\mathrm{s}}{\sqrt{F_{\rm IC}}} \;\;,
\end{equation*}
where $d_{\rm L} = 1.34$\,Gpc is the luminosity distance to \pks. From the values of $F_s$ and $F_{\rm IC}$ previously estimated for the energy fluxes of the peaks, this gives the constraint $\log_{10}(B\delta^3/1 \rm G) \geq 1.78$.

The SED in the low state is well reproduced by an SSC model with a set of parameters in agreement with the constraints derived above. The parameters are shown in Table~\ref{tab-ssc} where K is the normalisation factor of the electron spectrum, and $n_1$ and $n_2$ are the two spectral indices. The break energy $\gamma_\mathrm{b}$ and the energy range for the electrons $[\gamma_\mathrm{min}, \gamma_{\rm max}]$ are in units of $m_\mathrm{e}c^2$. A Doppler factor of 27 is found. This value is large enough to ensure that the IC emission fully takes place in the Thomson regime. The SSC parameters are in agreement with the observed variability timescale of 8 days found in the ATOM light curve that constrains the emitting region radius to $R_\mathrm{b} < 4.3 \times10^{17} \; \rm cm$. With these parameters, variability down to a timescale of 2 days can be theoretically produced. Although the emitting region radius of the model is small enough to be compatible with the minimal variability timescale measured, its value is one order of magnitude larger than what is commonly derived from one-zone SSC modelling of blazar SEDs \citep[see e.g.][]{2009PhDT.........5L,2010MNRAS.401.1570T}. Regarding the electron spectrum, the primary slope is hard compared to the canonical case $n = 2$ of standard Fermi-type acceleration mechanisms. Such a hard slope can theoretically be attained in some models of relativistic diffusive shock acceleration \citep[see e.g.][]{2012ApJ...745...63S}. Another limitation of the model is the large spectral break in the electron spectrum which cannot be associated to an equilibrium between the cooling and the escape of the electrons. In this scenario, a break value of one is expected at the electron Lorentz factor  $\left[\frac{4}{3}\frac{\sigma_{\rm T}}{m_{\rm e}c^2}R_{\rm b}U_{\rm B}\right]^{-1} \sim 10^5$~\citep{1998ApJ...509..608T} which is far beyond the value found for \pks. Similarly to other HBL \citep[see e.g.][]{2007ApJ...667L..21A,2010A+A...521A..69A,2010ApJ...715L..49A}, the model is far out of equipartition with a ratio of electron kinetic energy over magnetic energy of $U_{\rm e}/U_{\rm B}\sim$ 51 \citep[but see also][]{2013ApJ...771L...4C}.

Comparing the photon index of $\Gamma \simeq 1.94$ in the quiescent state at HE and $\Gamma \simeq 4.6$ for the \hess spectrum, the high energy part of the SED of \pks suggests the presence of a break between the HE and VHE regimes with a $\Delta\Gamma = 2.7 \pm 0.7$. Correcting the \hess spectrum for EBL absorption using the model of \cite{2008A+A...487..837F} leads to an index of $3.1 \pm 0.7$. Hence, the break stems from both a curvature of the intrinsic emitted spectrum and the absorption of the VHE spectrum from the EBL.

The synchrotron component  can  fully account for the optical and X-ray band emission. However, the model cannot explain infrared data and radio data, that are not contemporaneous to the low state. To account for the low infrared data, a primary slope for the electron spectrum as hard as 1 is needed. However, such a hard slope is very difficult to achieve in standard particle acceleration models. These data may therefore indicate that the source has been measured in a lower state. For radio data, the emission from the compact emitting region is self-absorbed and the interpretation requires a more extended emission zone in which the density of particles would be low enough to prevent self-absorption \citep{1989MNRAS.237..411S}. Indeed, radio observations of \pks at the kilo-parsec scale exhibit an extended diffuse halo-like component resolved around a bright core \citep{1998ApJS..118..327K}. 

As demonstrated in Sect.~\ref{subsec-atom}, a lack of significant correlation between the HE range and the ATOM data on the one hand, but an indication for correlated behaviour between the VHE and ATOM data on the other hand, were found. This would point to the fact that, in the framework of an SSC interpretation of the SED of \pks, the electrons radiating in optical and VHE would stem from the same underlying population, whereas a different population would be responsible for the HE emission. This behaviour was also detected in \object{PKS\,2155$-$304} in a low state of activity \citep{2009ApJ...696L.150A}. The lack of optical/HE correlation suggests that a simple one-zone SSC model is too simplistic to fully account for the time-dependent behaviour of \pks, since such a correlation is actually expected in these models. Multi-zone SSC models can account for extreme behaviour such as the so-called orphan \g-ray flare observed in \object{1ES\,1959$+$650} in 2002 \citep{2004ApJ...601..151K}, and can give an alternative explanation for this lack of optical/HE correlation.

\begin{table*}
\caption{SSC parameters used for the modelling of low state data (EBL absorption taken into account) and the ratio of kinetic energy density over magnetic energy density.}
\label{tab-ssc}
\centering
\begin{tabular}{ccccccccccc}
\hline\hline
 & B (mG)  & $\delta$ & $R_\mathrm{b}$ ($10^{16} \;$ cm)  & K (cm$^{-3}$) &  $n_1$ & $n_2$ & $\gamma_\mathrm{min}$ & $\gamma_\mathrm{b}$ & $\gamma_{\rm max}$ & $U_{\rm e}/U_{\rm B}$\\
\hline
Low state & 20 & 27 & 13 & 40 & 1.86 & 3.7 & 1 & 2.4$\times10^4$ & 6.9$\times10^5$ & 51 \\[3pt]
\hline
\end{tabular}
\end{table*}

\subsection{Constraints on the EBL}
\label{subsec-ebl}

The EBL is a diffuse extragalactic background of photons in the IR-UV bands. It stems from the light that has been emitted by galaxies through the history of the Universe, and part of which has been absorbed by interstellar dust and re-emitted in the infrared. Direct measurements of this diffuse component are contaminated by the bright foreground component associated with zodiacal light \citep{1998ApJ...508...25H} and models remain subject to large uncertainties. Extragalactic \g-ray source spectra are affected by absorption on this background light through the leptonic pair creation process (see \citealt{1967PhRv..155.1408G} or \citealt{2013APh....43..112D} for a recent review). The expected imprints of EBL absorption in blazar spectra can therefore be searched to put constraints on the level of EBL \citep{1992ApJ...390L..49S}. In this context, the advent of the \fermi instrument allows the intrinsic spectrum to be better constrained, which in turn improves the derived limit on the EBL \citep{2010ApJ...714L.157G,2011ApJ...733...77O,2012A+A...542A..59M}. Detecting second order effects in the brightest blazar spectra allows the level of the EBL to be measured at low redshifts $z < 0.2$ using IACTs like \hess \citep{2013A+A...550A...4H} and at higher redshifts $0.5 < z < 1.6$ with LAT observations \citep{2012Sci...338.1190A}. At first order, \g-ray absorption on the EBL in the 0.1-1 TeV range results in a steepening of the spectral index. One method to put constraints on the level of EBL is to assume a minimal slope for the intrinsic VHE spectrum and, by relating that to the measured slope, to estimate the maximum level of EBL allowed by this assumption. A reasonable and classical assumption for the minimal slope is 1.5 which is the hardest spectral index obtained for accelerated particles in non-relativistic shock acceleration models \citep[see e.g.][]{2001RPPh...64..429M}. Nevertheless, as discussed above, the choice of the minimal slope can be improved by using the LAT spectrum. Under the assumption that the slope at VHE is steeper than the slope at HE,  the minimal slope can be obtained from the \fermi spectrum taking into account its uncertainty. At the 3$\sigma$ confidence level, the minimal slope for this study is thus 1.86.

\begin{figure}
  \centering
  \includegraphics[width=\columnwidth]{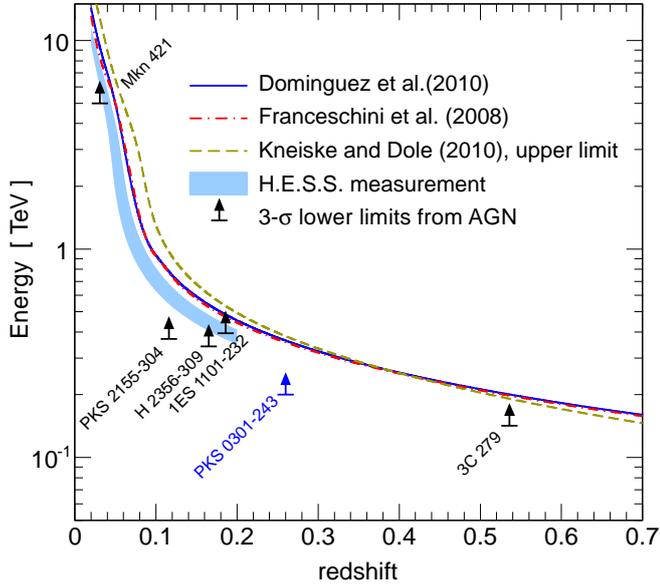}
  \caption{\g-ray horizon for two EBL models and the upper limit model from \cite{2010A+A...515A..19K}. The blue shaded area is the measurement by \hess from \cite{2013A+A...550A...4H} at 1$\sigma$ C.L. Some lower limits from AGN spectra measurements are shown, given at the 3$\sigma$ C.L.}
  \label{fig:ebl}
\end{figure}

To determine the upper limit on the level of EBL, the shape of the SED of the EBL is taken as in \citet{2008A+A...487..837F}. To ensure that the de-absorbed spectral slope is not less than the minimal slope assumed as 1.86, the level of EBL cannot be more than 2.7 times this EBL template at a confidence level (C.L.) of 99\%. This upper limit can be more explicitly expressed by defining the  \g-ray horizon \citep{1970Natur.226..135F} as the redshift giving an optical depth $\tau = 1$ to photons of a given energy.  This way, considering a fixed redshift $z$, the higher the level of EBL, the lower the energy needed to reach $\tau = 1$, so that an upper limit on the level of EBL translates to a lower limit on the energy of photons having a $\gamma$-ray horizon $z$. The constraint obtained from the spectrum of \pks thus yields a lower limit of 200 GeV on the energy for a \g-ray horizon at $z$ = 0.266 at the 3$\sigma$ level. Figure \ref{fig:ebl} shows the \g-ray horizon for two different EBL models \citep{2008A+A...487..837F,2011MNRAS.410.2556D} and one model considered to be a lower limit on the density of the EBL \citep{2010A+A...515A..19K}, thus translated to an upper limit on the figure. Also shown is the lower limit derived in this work and some other lower limits from AGN spectral measurements \citep{2006Natur.440.1018A,2008Sci...320.1752M,2005A+A...437...95A,2005A+A...430..865A}. The \hess measurement of the EBL density \citep{2013A+A...550A...4H} derived using a large sample of blazars is also shown.  All the lower limits from AGN measurements are compatible with the models, but one can see that the space allowed for models in this plane between the constraints is rather small.

\section{Conclusion}
\label{sec-conclusion}

The HBL \pks has been discovered in the VHE band with \hess The VHE emission shows a steep spectrum and no significant variability is detected. Pronounced variability is observed at HE with a strong \fermi flare during April-May 2010, and to a lesser extent in optical and X-rays. The minimal variability timescale of about 8 days observed in the optical band puts a limit on the size of the emitting region well within the range of observations of other HBL objects \citep[see e.g.][]{2010MNRAS.405L..94T}. An indication for a correlation between \hess and the optical ATOM data is found while no correlation is observed between LAT and ATOM, suggesting that an SSC model with one single zone may be too simplistic to explain the observed variability.

The SED of \pks shows a peak of the synchrotron component located in the UV band, a rather low frequency for an HBL object. Assuming a simple one-zone SSC model to interpret the emission from the optical to the VHE bands, the properties of the emitting region and the underlying particle energy distribution are investigated to interpret the low state of activity observed with \swift, \hess, and \fermi. A good agreement is found for a rather large and low density emitting region. The IC emission takes place in the Thomson regime and the model is dominated by the electron kinetic energy. The applicability of the one-zone SSC model to this data set is, however, limited by the lack of correlation between the optical and the HE domain.

The combined LAT and \hess data are also used to constrain the energy at which the Universe becomes opaque due to EBL absorption. It is found that the universe must become opaque ($\tau = 1$) to \g-rays from $z = 0.266$ at energies greater than 200 GeV in order to account for the spectral break observed between \fermi and \hess data.

\begin{acknowledgements}
  The support of the Namibian authorities and of the University of Namibia in facilitating the construction and operation of H.E.S.S. is gratefully acknowledged, as is the support by the German Ministry for Education and Research (BMBF), the Max Planck Society, the German Research Foundation (DFG), the French Ministry for Research, the CNRS-IN2P3 and the Astroparticle Interdisciplinary Programme of the CNRS, the U.K. Science and Technology Facilities Council (STFC), the IPNP of the Charles University, the Czech Science Foundation, the Polish Ministry of Science and  Higher Education, the South African Department of Science and Technology and National Research Foundation, and by the University of Namibia. We appreciate the excellent work of the technical support staff in Berlin, Durham, Hamburg, Heidelberg, Palaiseau, Paris, Saclay, and in Namibia in the construction and operation of the equipment.\\
  
  \indent The authors acknowledge the support by the \swift team for providing ToO observations, and the use of the public HEASARC software packages. This research made use of the NASA/IPAC Extragalactic Database (NED) and of the SIMBAD Astronomical Database. Part of this work is based on archival data, software, or on-line services provided by the ASI Science Data Center (ASDC).\\

  \indent The authors would like to thank the anonymous referee for the useful and constructive comments that contributed to improving this manuscript.
\end{acknowledgements}

\bibliographystyle{aa}  
\bibliography{pks0301_v6.0}

\begin{thebibliography}{103}
\expandafter\ifx\csname natexlab\endcsname\relax\def\natexlab#1{#1}\fi

\bibitem[{{Abdo} {et~al.}(2010{\natexlab{a}}){Abdo}, {Ackermann}, {Agudo},
  {Ajello}, {Aller}, {Aller}, {Angelakis}, {Arkharov}, {Axelsson}, {Bach},
  {Baldini}, {Ballet}, {Barbiellini}, {Bastieri}, {Baughman}, {Bechtol},
  {Bellazzini}, {Benitez}, {Berdyugin}, {Berenji}, {Blandford}, {Bloom},
  {Boettcher}, {Bonamente}, {Borgland}, {Bregeon}, {Brez}, {Brigida}, {Bruel},
  {Burnett}, {Burrows}, {Buson}, {Caliandro}, {Calzoletti}, {Cameron},
  {Capalbi}, {Caraveo}, {Carosati}, {Casandjian}, {Cavazzuti}, {Cecchi}, {{\c
  C}elik}, {Charles}, {Chaty}, {Chekhtman}, {Chen}, {Chiang}, {Chincarini},
  {Ciprini}, {Claus}, {Cohen-Tanugi}, {Colafrancesco}, {Cominsky}, {Conrad},
  {Costamante}, {Cutini}, {D'ammando}, {Deitrick}, {D'Elia}, {Dermer}, {de
  Angelis}, {de Palma}, {Digel}, {Donnarumma}, {Silva}, {Drell}, {Dubois},
  {Dultzin}, {Dumora}, {Falcone}, {Farnier}, {Favuzzi}, {Fegan}, {Focke},
  {Forn{\'e}}, {Fortin}, {Frailis}, {Fuhrmann}, {Fukazawa}, {Funk}, {Fusco},
  {G{\'o}mez}, {Gargano}, {Gasparrini}, {Gehrels}, {Germani}, {Giebels},
  {Giglietto}, {Giommi}, {Giordano}, {Giuliani}, {Glanzman}, {Godfrey},
  {Grenier}, {Gronwall}, {Grove}, {Guillemot}, {Guiriec}, {Gurwell}, {Hadasch},
  {Hanabata}, {Harding}, {Hayashida}, {Hays}, {Healey}, {Heidt}, {Hiriart},
  {Horan}, {Hoversten}, {Hughes}, {Itoh}, {Jackson}, {J{\'o}hannesson},
  {Johnson}, {Johnson}, {Jorstad}, {Kadler}, {Kamae}, {Katagiri}, {Kataoka},
  {Kawai}, {Kennea}, {Kerr}, {Kimeridze}, {Kn{\"o}dlseder}, {Kocian},
  {Kopatskaya}, {Koptelova}, {Konstantinova}, {Kovalev}, {Kovalev},
  {Kurtanidze}, {Kuss}, {Lande}, {Larionov}, {Latronico}, {Leto}, {Lindfors},
  {Longo}, {Loparco}, {Lott}, {Lovellette}, {Lubrano}, {Madejski}, {Makeev},
  {Marchegiani}, {Marscher}, {Marshall}, {Max-Moerbeck}, {Mazziotta},
  {McConville}, {McEnery}, {Meurer}, {Michelson}, {Mitthumsiri}, {Mizuno},
  {Moiseev}, {Monte}, {Monzani}, {Morselli}, {Moskalenko}, {Murgia},
  {Nestoras}, {Nilsson}, {Nizhelsky}, {Nolan}, {Norris}, {Nuss}, {Ohsugi},
  {Ojha}, {Omodei}, {Orlando}, {Ormes}, {Osborne}, {Ozaki}, {Pacciani},
  {Padovani}, {Pagani}, {Page}, {Paneque}, {Panetta}, {Parent}, {Pasanen},
  {Pavlidou}, {Pelassa}, {Pepe}, {Perri}, {Pesce-Rollins}, {Piranomonte},
  {Piron}, {Pittori}, {Porter}, {Puccetti}, {Rahoui}, {Rain{\`o}}, {Raiteri},
  {Rando}, {Razzano}, {Reimer}, {Reimer}, {Reposeur}, {Richards}, {Ritz},
  {Rochester}, {Rodriguez}, {Romani}, {Ros}, {Roth}, {Roustazadeh}, {Ryde},
  {Sadrozinski}, {Sadun}, {Sanchez}, {Sander}, {Saz Parkinson}, {Scargle},
  {Sellerholm}, {Sgr{\`o}}, {Shaw}, {Sigua}, {Siskind}, {Smith}, {Smith},
  {Spandre}, {Spinelli}, {Starck}, {Stevenson}, {Stratta}, {Strickman},
  {Suson}, {Tajima}, {Takahashi}, {Takahashi}, {Takalo}, {Tanaka}, {Thayer},
  {Thayer}, {Thompson}, {Tibaldo}, {Torres}, {Tosti}, {Tramacere}, {Uchiyama},
  {Usher}, {Vasileiou}, {Verrecchia}, {Vilchez}, {Villata}, {Vitale}, {Waite},
  {Wang}, {Winer}, {Wood}, {Ylinen}, {Zensus}, {Zhekanis}, \&
  {Ziegler}}]{2010ApJ...716...30A}
{Abdo}, A.~A., {Ackermann}, M., {Agudo}, I., {et~al.} (\fermi Collaboration) 2010{\natexlab{a}}, \apj,
  716, 30

\bibitem[{{Abdo} {et~al.}(2010{\natexlab{b}}){Abdo}, {Ackermann}, {Ajello},
  {Allafort}, {Antolini}, {Atwood}, {Axelsson}, {Baldini}, {Ballet},
  {Barbiellini}, {Bastieri}, {Baughman}, {Bechtol}, {Bellazzini}, {Belli},
  {Berenji}, {Bisello}, {Blandford}, {Bloom}, {Bonamente}, {Bonnell},
  {Borgland}, {Bouvier}, {Bregeon}, {Brez}, {Brigida}, {Bruel}, {Burnett},
  {Busetto}, {Buson}, {Caliandro}, {Cameron}, {Campana}, {Canadas}, {Caraveo},
  {Carrigan}, {Casandjian}, {Cavazzuti}, {Ceccanti}, {Cecchi}, {{\c C}elik},
  {Charles}, {Chekhtman}, {Cheung}, {Chiang}, {Cillis}, {Ciprini}, {Claus},
  {Cohen-Tanugi}, {Conrad}, {Corbet}, {Davis}, {DeKlotz}, {den Hartog},
  {Dermer}, {de Angelis}, {de Luca}, {de Palma}, {Digel}, {Dormody}, {Silva},
  {Drell}, {Dubois}, {Dumora}, {Fabiani}, {Farnier}, {Favuzzi}, {Fegan},
  {Ferrara}, {Focke}, {Fortin}, {Frailis}, {Fukazawa}, {Funk}, {Fusco},
  {Gargano}, {Gasparrini}, {Gehrels}, {Germani}, {Giavitto}, {Giebels},
  {Giglietto}, {Giommi}, {Giordano}, {Giroletti}, {Glanzman}, {Godfrey},
  {Grenier}, {Grondin}, {Grove}, {Guillemot}, {Guiriec}, {Gustafsson},
  {Hadasch}, {Hanabata}, {Harding}, {Hayashida}, {Hays}, {Healey}, {Hill},
  {Horan}, {Hughes}, {Iafrate}, {J{\'o}hannesson}, {Johnson}, {Johnson},
  {Johnson}, {Johnson}, {Kamae}, {Katagiri}, {Kataoka}, {Kawai}, {Kerr},
  {Kn{\"o}dlseder}, {Kocevski}, {Kuss}, {Lande}, {Landriu}, {Latronico}, {Lee},
  {Lemoine-Goumard}, {Lionetto}, {Llena Garde}, {Longo}, {Loparco}, {Lott},
  {Lovellette}, {Lubrano}, {Madejski}, {Makeev}, {Marangelli}, {Marelli},
  {Massaro}, {Mazziotta}, {McConville}, {McEnery}, {Michelson}, {Minuti},
  {Mitthumsiri}, {Mizuno}, {Moiseev}, {Mongelli}, {Monte}, {Monzani},
  {Moretti}, {Morselli}, {Moskalenko}, {Murgia}, {Nakajima}, {Nakamori},
  {Naumann-Godo}, {Nolan}, {Norris}, {Nuss}, {Ohno}, {Ohsugi}, {Omodei},
  {Orlando}, {Ormes}, {Ozaki}, {Paccagnella}, {Paneque}, {Panetta}, {Parent},
  {Pelassa}, {Pepe}, {Pesce-Rollins}, {Pinchera}, {Piron}, {Porter}, {Poupard},
  {Rain{\`o}}, {Rando}, {Ray}, {Razzano}, {Razzaque}, {Rea}, {Reimer},
  {Reimer}, {Reposeur}, {Ripken}, {Ritz}, {Rochester}, {Rodriguez}, {Romani},
  {Roth}, {Sadrozinski}, {Salvetti}, {Sanchez}, {Sander}, {Saz Parkinson},
  {Scargle}, {Schalk}, {Scolieri}, {Sgr{\`o}}, {Shaw}, {Siskind}, {Smith},
  {Smith}, {Spandre}, {Spinelli}, {Starck}, {Stephens}, {Striani}, {Strickman},
  {Strong}, {Suson}, {Tajima}, {Takahashi}, {Takahashi}, {Tanaka}, {Thayer},
  {Thayer}, {Thompson}, {Tibaldo}, {Tibolla}, {Tinebra}, {Torres}, {Tosti},
  {Tramacere}, {Uchiyama}, {Usher}, {Van Etten}, {Vasileiou}, {Vilchez},
  {Vitale}, {Waite}, {Wallace}, {Wang}, {Watters}, {Winer}, {Wood}, {Yang},
  {Ylinen}, \& {Ziegler}}]{2010ApJS..188..405A}
{Abdo}, A.~A., {Ackermann}, M., {Ajello}, M., {et~al.} (\fermi Collaboration) 2010{\natexlab{b}},
  \apjs, 188, 405

\bibitem[{{Abdo} {et~al.}(2010{\natexlab{c}}){Abdo}, {Ackermann}, {Ajello},
  {Allafort}, {Antolini}, {Atwood}, {Axelsson}, {Baldini}, {Ballet},
  {Barbiellini}, {Bastieri}, {Baughman}, {Bechtol}, {Bellazzini}, {Berenji},
  {Blandford}, {Bloom}, {Bogart}, {Bonamente}, {Borgland}, {Bouvier},
  {Bregeon}, {Brez}, {Brigida}, {Bruel}, {Buehler}, {Burnett}, {Buson},
  {Caliandro}, {Cameron}, {Cannon}, {Caraveo}, {Carrigan}, {Casandjian},
  {Cavazzuti}, {Cecchi}, {{\c C}elik}, {Celotti}, {Charles}, {Chekhtman},
  {Chen}, {Cheung}, {Chiang}, {Ciprini}, {Claus}, {Cohen-Tanugi}, {Conrad},
  {Costamante}, {Cotter}, {Cutini}, {D'Elia}, {Dermer}, {de Angelis}, {de
  Palma}, {De Rosa}, {Digel}, {Silva}, {Drell}, {Dubois}, {Dumora}, {Escande},
  {Farnier}, {Favuzzi}, {Fegan}, {Ferrara}, {Focke}, {Fortin}, {Frailis},
  {Fukazawa}, {Funk}, {Fusco}, {Gargano}, {Gasparrini}, {Gehrels}, {Germani},
  {Giebels}, {Giglietto}, {Giommi}, {Giordano}, {Giroletti}, {Glanzman},
  {Godfrey}, {Grandi}, {Grenier}, {Grondin}, {Grove}, {Guiriec}, {Hadasch},
  {Harding}, {Hayashida}, {Hays}, {Healey}, {Hill}, {Horan}, {Hughes},
  {Iafrate}, {Itoh}, {J{\'o}hannesson}, {Johnson}, {Johnson}, {Johnson},
  {Johnson}, {Kamae}, {Katagiri}, {Kataoka}, {Kawai}, {Kerr}, {Kn{\"o}dlseder},
  {Kuss}, {Lande}, {Latronico}, {Lavalley}, {Lemoine-Goumard}, {Llena Garde},
  {Longo}, {Loparco}, {Lott}, {Lovellette}, {Lubrano}, {Madejski}, {Makeev},
  {Malaguti}, {Massaro}, {Mazziotta}, {McConville}, {McEnery}, {McGlynn},
  {Michelson}, {Mitthumsiri}, {Mizuno}, {Moiseev}, {Monte}, {Monzani},
  {Morselli}, {Moskalenko}, {Murgia}, {Nolan}, {Norris}, {Nuss}, {Ohno},
  {Ohsugi}, {Omodei}, {Orlando}, {Ormes}, {Ozaki}, {Paneque}, {Panetta},
  {Parent}, {Pelassa}, {Pepe}, {Pesce-Rollins}, {Piranomonte}, {Piron},
  {Porter}, {Rain{\`o}}, {Rando}, {Razzano}, {Reimer}, {Reimer}, {Reposeur},
  {Ripken}, {Ritz}, {Rodriguez}, {Romani}, {Roth}, {Ryde}, {Sadrozinski},
  {Sanchez}, {Sander}, {Saz Parkinson}, {Scargle}, {Sgr{\`o}}, {Shaw},
  {Siskind}, {Smith}, {Spandre}, {Spinelli}, {Starck}, {Stawarz}, {Strickman},
  {Suson}, {Tajima}, {Takahashi}, {Takahashi}, {Tanaka}, {Taylor}, {Thayer},
  {Thayer}, {Thompson}, {Tibaldo}, {Torres}, {Tosti}, {Tramacere}, {Ubertini},
  {Uchiyama}, {Usher}, {Vasileiou}, {Vilchez}, {Villata}, {Vitale}, {Waite},
  {Wallace}, {Wang}, {Winer}, {Wood}, {Yang}, {Ylinen}, \&
  {Ziegler}}]{2010ApJ...715..429A}
{Abdo}, A.~A., {Ackermann}, M., {Ajello}, M., {et~al.} (\fermi Collaboration) 2010{\natexlab{c}},
  \apj, 715, 429

\bibitem[{{Abdo} {et~al.}(2009){Abdo}, {Ackermann}, {Ajello}, {Atwood},
  {Axelsson}, {Baldini}, {Ballet}, {Band}, {Barbiellini}, {Bastieri},
  {Battelino}, {Baughman}, {Bechtol}, {Bellazzini}, {Berenji}, {Bignami},
  {Blandford}, {Bloom}, {Bonamente}, {Borgland}, {Bouvier}, {Bregeon}, {Brez},
  {Brigida}, {Bruel}, {Burnett}, {Caliandro}, {Cameron}, {Caraveo},
  {Casandjian}, {Cavazzuti}, {Cecchi}, {Charles}, {Chekhtman}, {Cheung},
  {Chiang}, {Ciprini}, {Claus}, {Cohen-Tanugi}, {Cominsky}, {Conrad}, {Corbet},
  {Costamante}, {Cutini}, {Davis}, {Dermer}, {de Angelis}, {de Luca}, {de
  Palma}, {Digel}, {Dormody}, {do Couto e Silva}, {Drell}, {Dubois}, {Dumora},
  {Farnier}, {Favuzzi}, {Fegan}, {Ferrara}, {Focke}, {Frailis}, {Fukazawa},
  {Funk}, {Fusco}, {Gargano}, {Gasparrini}, {Gehrels}, {Germani}, {Giebels},
  {Giglietto}, {Giommi}, {Giordano}, {Glanzman}, {Godfrey}, {Grenier},
  {Grondin}, {Grove}, {Guillemot}, {Guiriec}, {Hanabata}, {Harding}, {Hartman},
  {Hayashida}, {Hays}, {Healey}, {Horan}, {Hughes}, {J{\'o}hannesson},
  {Johnson}, {Johnson}, {Johnson}, {Johnson}, {Kamae}, {Katagiri}, {Kataoka},
  {Kawai}, {Kerr}, {Kn{\"o}dlseder}, {Kocevski}, {Kocian}, {Komin}, {Kuehn},
  {Kuss}, {Lande}, {Latronico}, {Lee}, {Lemoine-Goumard}, {Longo}, {Loparco},
  {Lott}, {Lovellette}, {Lubrano}, {Madejski}, {Makeev}, {Marelli},
  {Mazziotta}, {McConville}, {McEnery}, {McGlynn}, {Meurer}, {Michelson},
  {Mitthumsiri}, {Mizuno}, {Moiseev}, {Monte}, {Monzani}, {Moretti},
  {Morselli}, {Moskalenko}, {Murgia}, {Nakamori}, {Nolan}, {Norris}, {Nuss},
  {Ohno}, {Ohsugi}, {Omodei}, {Orlando}, {Ormes}, {Ozaki}, {Paneque},
  {Panetta}, {Parent}, {Pelassa}, {Pepe}, {Pesce-Rollins}, {Piron}, {Porter},
  {Poupard}, {Rain{\`o}}, {Rando}, {Ray}, {Razzano}, {Rea}, {Reimer}, {Reimer},
  {Reposeur}, {Ritz}, {Rochester}, {Rodriguez}, {Romani}, {Roth}, {Ryde},
  {Sadrozinski}, {Sanchez}, {Sander}, {Saz Parkinson}, {Scargle}, {Schalk},
  {Sellerholm}, {Sgr{\`o}}, {Shaw}, {Shrader}, {Sierpowska-Bartosik},
  {Siskind}, {Smith}, {Smith}, {Spandre}, {Spinelli}, {Starck}, {Stephens},
  {Strickman}, {Strong}, {Suson}, {Tajima}, {Takahashi}, {Takahashi}, {Tanaka},
  {Thayer}, {Thayer}, {Thompson}, {Tibaldo}, {Tibolla}, {Torres}, {Tosti},
  {Tramacere}, {Uchiyama}, {Usher}, {Van Etten}, {Vilchez}, {Vitale}, {Waite},
  {Wallace}, {Wang}, {Watters}, {Winer}, {Wood}, {Ylinen}, {Ziegler}, \& {The
  Fermi/LAT Collaboration}}]{2009ApJS..183...46A}
{Abdo}, A.~A., {Ackermann}, M., {Ajello}, M., {et~al.} (\fermi Collaboration) 2009, \apjs, 183, 46

\bibitem[{{Abramowski} {et~al.}(2012){Abramowski}, {Acero}, {Aharonian},
  {Akhperjanian}, {Anton}, {Balzer}, {Barnacka}, {Barres de Almeida},
  {Becherini}, {Becker}, \& et~al.}]{2012ApJ...746..151A}
{Abramowski}, A., {Acero}, F., {Aharonian}, F., {et~al.} (\hess, MAGIC, VERITAS, \fermi Collaborations) 2012, \apj, 746, 151

\bibitem[{{Acciari} {et~al.}(2010){Acciari}, {Aliu}, {Arlen}, {Aune},
  {Bautista}, {Beilicke}, {Benbow}, {B{\"o}ttcher}, {Boltuch}, {Bradbury},
  {Buckley}, {Bugaev}, {Byrum}, {Cannon}, {Cesarini}, {Ciupik}, {Cui},
  {Dickherber}, {Duke}, {Falcone}, {Finley}, {Finnegan}, {Fortson}, {Furniss},
  {Galante}, {Gall}, {Gibbs}, {Gillanders}, {Godambe}, {Grube}, {Guenette},
  {Gyuk}, {Hanna}, {Holder}, {Hui}, {Humensky}, {Imran}, {Kaaret}, {Karlsson},
  {Kertzman}, {Kieda}, {Konopelko}, {Krawczynski}, {Krennrich}, {Lang},
  {Lamerato}, {LeBohec}, {Maier}, {McArthur}, {McCann}, {McCutcheon},
  {Moriarty}, {Mukherjee}, {Ong}, {Otte}, {Pandel}, {Perkins}, {Petry},
  {Pichel}, {Pohl}, {Quinn}, {Ragan}, {Reyes}, {Reynolds}, {Roache}, {Rose},
  {Roustazadeh}, {Schroedter}, {Sembroski}, {Senturk}, {Smith}, {Steele},
  {Swordy}, {Te{\v s}i{\'c}}, {Theiling}, {Thibadeau}, {Varlotta}, {Vassiliev},
  {Vincent}, {Wagner}, {Wakely}, {Ward}, {Weekes}, {Weinstein}, {Weisgarber},
  {Williams}, {Wissel}, {Wood}, {Zitzer}, {Ackermann}, {Ajello}, {Antolini},
  {Baldini}, {Ballet}, {Barbiellini}, {Bastieri}, {Bechtol}, {Bellazzini},
  {Berenji}, {Blandford}, {Bloom}, {Bonamente}, {Borgland}, {Bouvier},
  {Bregeon}, {Brigida}, {Bruel}, {Buehler}, {Buson}, {Caliandro}, {Cameron},
  {Caraveo}, {Carrigan}, {Casandjian}, {Cavazzuti}, {Cecchi}, {{\c C}elik},
  {Charles}, {Chekhtman}, {Cheung}, {Chiang}, {Ciprini}, {Claus},
  {Cohen-Tanugi}, {Conrad}, {Dermer}, {de Palma}, {Silva}, {Drell}, {Dubois},
  {Dumora}, {Farnier}, {Favuzzi}, {Fegan}, {Fortin}, {Frailis}, {Fukazawa},
  {Funk}, {Fusco}, {Gargano}, {Gasparrini}, {Gehrels}, {Germani}, {Giebels},
  {Giglietto}, {Giordano}, {Giroletti}, {Glanzman}, {Godfrey}, {Grenier},
  {Grove}, {Guiriec}, {Hays}, {Horan}, {Hughes}, {J{\'o}hannesson}, {Johnson},
  {Johnson}, {Kamae}, {Katagiri}, {Kataoka}, {Kn{\"o}dlseder}, {Kuss}, {Lande},
  {Latronico}, {Lee}, {Llena Garde}, {Longo}, {Loparco}, {Lott}, {Lovellette},
  {Lubrano}, {Makeev}, {Mazziotta}, {Michelson}, {Mitthumsiri}, {Mizuno},
  {Moiseev}, {Monte}, {Monzani}, {Morselli}, {Moskalenko}, {Murgia}, {Nolan},
  {Norris}, {Nuss}, {Ohno}, {Ohsugi}, {Omodei}, {Orlando}, {Ormes}, {Paneque},
  {Panetta}, {Pelassa}, {Pepe}, {Pesce-Rollins}, {Piron}, {Porter},
  {Rain{\`o}}, {Rando}, {Razzano}, {Reimer}, {Reimer}, {Ripken}, {Rodriguez},
  {Roth}, {Sadrozinski}, {Sanchez}, {Sander}, {Scargle}, {Sgr{\`o}}, {Siskind},
  {Smith}, {Spandre}, {Spinelli}, {Strickman}, {Suson}, {Takahashi}, {Tanaka},
  {Thayer}, {Thayer}, {Thompson}, {Tibaldo}, {Torres}, {Tosti}, {Tramacere},
  {Usher}, {Vasileiou}, {Vilchez}, {Vitale}, {Waite}, {Wang}, {Winer}, {Wood},
  {Yang}, {Ylinen}, \& {Ziegler}}]{2010ApJ...715L..49A}
{Acciari}, V.~A., {Aliu}, E., {Arlen}, T., {et~al.} (VERITAS Collaboration) 2010, \apjl, 715, L49

\bibitem[{{Ackermann} {et~al.}(2012){Ackermann}, {Ajello}, {Allafort},
  {Schady}, {Baldini}, {Ballet}, {Barbiellini}, {Bastieri}, {Bellazzini},
  {Blandford}, {Bloom}, {Borgland}, {Bottacini}, {Bouvier}, {Bregeon},
  {Brigida}, {Bruel}, {Buehler}, {Buson}, {Caliandro}, {Cameron}, {Caraveo},
  {Cavazzuti}, {Cecchi}, {Charles}, {Chaves}, {Chekhtman}, {Cheung}, {Chiang},
  {Chiaro}, {Ciprini}, {Claus}, {Cohen-Tanugi}, {Conrad}, {Cutini},
  {D'Ammando}, {de Palma}, {Dermer}, {Digel}, {do Couto e Silva},
  {Dom{\'{\i}}nguez}, {Drell}, {Drlica-Wagner}, {Favuzzi}, {Fegan}, {Focke},
  {Franckowiak}, {Fukazawa}, {Funk}, {Fusco}, {Gargano}, {Gasparrini},
  {Gehrels}, {Germani}, {Giglietto}, {Giordano}, {Giroletti}, {Glanzman},
  {Godfrey}, {Grenier}, {Grove}, {Guiriec}, {Gustafsson}, {Hadasch},
  {Hayashida}, {Hays}, {Jackson}, {Jogler}, {Kataoka}, {Kn{\"o}dlseder},
  {Kuss}, {Lande}, {Larsson}, {Latronico}, {Longo}, {Loparco}, {Lovellette},
  {Lubrano}, {Mazziotta}, {McEnery}, {Mehault}, {Michelson}, {Mizuno}, {Monte},
  {Monzani}, {Morselli}, {Moskalenko}, {Murgia}, {Tramacere}, {Nuss},
  {Greiner}, {Ohno}, {Ohsugi}, {Omodei}, {Orienti}, {Orlando}, {Ormes},
  {Paneque}, {Perkins}, {Pesce-Rollins}, {Piron}, {Pivato}, {Porter},
  {Rain{\`o}}, {Rando}, {Razzano}, {Razzaque}, {Reimer}, {Reimer}, {Reyes},
  {Ritz}, {Rau}, {Romoli}, {Roth}, {S{\'a}nchez-Conde}, {Sanchez}, {Scargle},
  {Sgr{\`o}}, {Siskind}, {Spandre}, {Spinelli}, {Stawarz}, {Suson},
  {Takahashi}, {Tanaka}, {Thayer}, {Thompson}, {Tibaldo}, {Tinivella},
  {Torres}, {Tosti}, {Troja}, {Usher}, {Vandenbroucke}, {Vasileiou},
  {Vianello}, {Vitale}, {Waite}, {Winer}, {Wood}, \&
  {Wood}}]{2012Sci...338.1190A}
{Ackermann}, M., {Ajello}, M., {Allafort}, A., {et~al.} (\fermi Collaboration) 2012, Science, 338,
  1190

\bibitem[{{Aharonian} {et~al.}(2010){Aharonian}, {Akhperjanian}, {Anton},
  {Barres de Almeida}, {Bazer-Bachi}, {Becherini}, {Behera}, {Benbow},
  {Bernl{\"o}hr}, {Bochow}, {Boisson}, {Bolmont}, {Borrel}, {Brucker}, {Brun},
  {Brun}, {B{\"u}hler}, {Bulik}, {B{\"u}sching}, {Boutelier}, {Chadwick},
  {Charbonnier}, {Chaves}, {Cheesebrough}, {Chounet}, {Clapson}, {Coignet},
  {Dalton}, {Daniel}, {Davids}, {Degrange}, {Deil}, {Dickinson},
  {Djannati-Ata{\"i}}, {Domainko}, {O'C.~Drury}, {Dubois}, {Dubus}, {Dyks},
  {Dyrda}, {Egberts}, {Emmanoulopoulos}, {Espigat}, {Farnier}, {Feinstein},
  {Fiasson}, {F{\"o}rster}, {Fontaine}, {F{\"u}{\ss}ling}, {Gabici}, {Gallant},
  {G{\'e}rard}, {Gerbig}, {Giebels}, {Glicenstein}, {Gl{\"u}ck}, {Goret},
  {G{\"o}ring}, {Hauser}, {Hauser}, {Heinz}, {Heinzelmann}, {Henri}, {Hermann},
  {Hinton}, {Hoffmann}, {Hofmann}, {Holleran}, {Hoppe}, {Horns},
  {Jacholkowska}, {de Jager}, {Jahn}, {Jung}, {Katarzy{\'n}ski}, {Katz},
  {Kaufmann}, {Kendziorra}, {Kerschhaggl}, {Khangulyan}, {Kh{\'e}lifi},
  {Keogh}, {Klu{\'z}niak}, {Kneiske}, {Komin}, {Kosack}, {Lamanna}, {Lenain},
  {Lohse}, {Marandon}, {Martin}, {Martineau-Huynh}, {Marcowith}, {Masbou},
  {Maurin}, {McComb}, {Medina}, {Moderski}, {Moulin}, {Naumann-Godo}, {de
  Naurois}, {Nedbal}, {Nekrassov}, {Nicholas}, {Niemiec}, {Nolan}, {Ohm},
  {Olive}, {de O{\~n}a Wilhelmi}, {Orford}, {Ostrowski}, {Panter}, {Paz
  Arribas}, {Pedaletti}, {Pelletier}, {Petrucci}, {Pita}, {P{\"u}hlhofer},
  {Punch}, {Quirrenbach}, {Raubenheimer}, {Raue}, {Rayner}, {Renaud}, {Rieger},
  {Ripken}, {Rob}, {Rosier-Lees}, {Rowell}, {Rudak}, {Rulten}, {Ruppel},
  {Sahakian}, {Santangelo}, {Schlickeiser}, {Sch{\"o}ck}, {Schr{\"o}der},
  {Schwanke}, {Schwarzburg}, {Schwemmer}, {Shalchi}, {Sikora}, {Skilton},
  {Sol}, {Spangler}, {Stawarz}, {Steenkamp}, {Stegmann}, {Stinzing},
  {Superina}, {Szostek}, {Tam}, {Tavernet}, {Terrier}, {Tibolla}, {Tluczykont},
  {van Eldik}, {Vasileiadis}, {Venter}, {Venter}, {Vialle}, {Vincent},
  {Vivier}, {V{\"o}lk}, {Volpe}, {Wagner}, {Ward}, {Zdziarski}, \&
  {Zech}}]{2010A+A...521A..69A}
{Aharonian}, F., {Akhperjanian}, A.~G., {Anton}, G., {et~al.} (\hess Collaboration) 2010, \aap, 521,
  A69+

\bibitem[{{Aharonian} {et~al.}(2009{\natexlab{a}}){Aharonian}, {Akhperjanian},
  {Anton}, {Barres de Almeida}, {Bazer-Bachi}, {Becherini}, {Behera},
  {Bernl{\"o}hr}, {Boisson}, {Bochow}, {Borrel}, {Brion}, {Brucker}, {Brun},
  {B{\"u}hler}, {Bulik}, {B{\"u}sching}, {Boutelier}, {Chadwick},
  {Charbonnier}, {Chaves}, {Cheesebrough}, {Chounet}, {Clapson}, {Coignet},
  {Dalton}, {Daniel}, {Davids}, {Degrange}, {Deil}, {Dickinson},
  {Djannati-Ata{\"i}}, {Domainko}, {O'C.~Drury}, {Dubois}, {Dubus}, {Dyks},
  {Dyrda}, {Egberts}, {Emmanoulopoulos}, {Espigat}, {Farnier}, {Feinstein},
  {Fiasson}, {F{\"o}rster}, {Fontaine}, {F{\"u}{\ss}ling}, {Gabici}, {Gallant},
  {G{\'e}rard}, {Giebels}, {Glicenstein}, {Gl{\"u}ck}, {Goret}, {G{\"o}hring},
  {Hauser}, {Hauser}, {Heinz}, {Heinzelmann}, {Henri}, {Hermann}, {Hinton},
  {Hoffmann}, {Hofmann}, {Holleran}, {Hoppe}, {Horns}, {Jacholkowska}, {de
  Jager}, {Jahn}, {Jung}, {Katarzy{\'n}ski}, {Katz}, {Kaufmann}, {Kendziorra},
  {Kerschhaggl}, {Khangulyan}, {Kh{\'e}lifi}, {Keogh}, {Klu{\'z}niak}, {Komin},
  {Kosack}, {Lamanna}, {Lenain}, {Lohse}, {Marandon}, {Martin},
  {Martineau-Huynh}, {Marcowith}, {Maurin}, {McComb}, {Medina}, {Moderski},
  {Moulin}, {Naumann-Godo}, {de Naurois}, {Nedbal}, {Nekrassov}, {Niemiec},
  {Nolan}, {Ohm}, {Olive}, {de O{\~n}a Wilhelmi}, {Orford}, {Ostrowski},
  {Panter}, {Arribas}, {Pedaletti}, {Pelletier}, {Petrucci}, {Pita},
  {P{\"u}hlhofer}, {Punch}, {Quirrenbach}, {Raubenheimer}, {Raue}, {Rayner},
  {Renaud}, {Rieger}, {Ripken}, {Rob}, {Rosier-Lees}, {Rowell}, {Rudak},
  {Rulten}, {Ruppel}, {Sahakian}, {Santangelo}, {Schlickeiser}, {Sch{\"o}ck},
  {Schr{\"o}der}, {Schwanke}, {Schwarzburg}, {Schwemmer}, {Shalchi}, {Sikora},
  {Skilton}, {Sol}, {Spangler}, {Stawarz}, {Steenkamp}, {Stegmann}, {Superina},
  {Szostek}, {Tam}, {Tavernet}, {Terrier}, {Tibolla}, {van Eldik},
  {Vasileiadis}, {Venter}, {Venter}, {Vialle}, {Vincent}, {Vivier}, {V{\"o}lk},
  {Volpe}, {Wagner}, {Ward}, {Zdziarski}, {Zech}, {Abdo}, {Ackermann},
  {Ajello}, {Atwood}, {Axelsson}, {Baldini}, {Ballet}, {Barbiellini}, {Baring},
  {Bastieri}, {Battelino}, {Baughman}, {Bechtol}, {Bellazzini}, {Berenji},
  {Bloom}, {Bonamente}, {Borgland}, {Bregeon}, {Brez}, {Brigida}, {Bruel},
  {Caliandro}, {Cameron}, {Caraveo}, {Casandjian}, {Cavazzuti}, {Cecchi},
  {Charles}, {Chekhtman}, {Chen}, {Cheung}, {Chiang}, {Ciprini}, {Claus},
  {Cohen-Tanugi}, {Colafrancesco}, {Conrad}, {Costamante}, {Cutini}, {Dermer},
  {de Angelis}, {de Palma}, {Digel}, {do Couto e Silva}, {Drell}, {Dubois},
  {Dubus}, {Dumora}, {Farnier}, {Favuzzi}, {Fegan}, {Ferrara}, {Fleury},
  {Focke}, {Frailis}, {Fukazawa}, {Funk}, {Fusco}, {Gargano}, {Gasparrini},
  {Gehrels}, {Germani}, {Giebels}, {Giglietto}, {Giordano}, {Grondin}, {Grove},
  {Guillemot}, {Guiriec}, {Hanabata}, {Harding}, {Hayashida}, {Hays}, {Horan},
  {J{\'o}hannesson}, {Johnson}, {Johnson}, {Johnson}, {Kadler}, {Kamae},
  {Katagiri}, {Kataoka}, {Kerr}, {Kn{\"o}dlseder}, {Kuehn}, {Kuss}, {Lande},
  {Latronico}, {Lee}, {Lemoine-Goumard}, {Longo}, {Loparco}, {Lott},
  {Lovellette}, {Madejski}, {Makeev}, {Mazziotta}, {McEnery}, {Meurer},
  {Michelson}, {Mitthumsiri}, {Mizuno}, {Moiseev}, {Monte}, {Monzani},
  {Morselli}, {Moskalenko}, {Murgia}, {Nolan}, {Nuss}, {Ohsugi}, {Omodei},
  {Orlando}, {Ormes}, {Paneque}, {Panetta}, {Parent}, {Pelassa}, {Pepe},
  {Pesce-Rollins}, {Piron}, {Porter}, {Rain{\`o}}, {Razzano}, {Reimer},
  {Reimer}, {Reposeur}, {Ritz}, {Rodriguez}, {Ryde}, {Sadrozinski}, {Sanchez},
  {Sander}, {Scargle}, {Schalk}, {Sellerholm}, {Sgr{\`o}}, {Shaw}, {Smith},
  {Spandre}, {Spinelli}, {Starck}, {Strickman}, {Tajima}, {Takahashi},
  {Takahashi}, {Tanaka}, {Thayer}, {Thompson}, {Tibaldo}, {Torres}, {Tosti},
  {Tramacere}, {Uchiyama}, {Usher}, {Vilchez}, {Villata}, {Vitale}, {Waite},
  {Wood}, {Ylinen}, \& {Ziegler}}]{2009ApJ...696L.150A}
{Aharonian}, F., {Akhperjanian}, A.~G., {Anton}, G., {et~al.} (\hess, \fermi Collaborations)
  2009{\natexlab{a}}, \apjl, 696, L150

\bibitem[{{Aharonian} {et~al.}(2009{\natexlab{b}}){Aharonian}, {Akhperjanian},
  {Anton}, {de Almeida}, {Bazer-Bachi}, {Becherini}, {Behera}, {Benbow},
  {Bernl{\"o}hr}, {Boisson}, {Bochow}, {Borrel}, {Brion}, {Brucker}, {Brun},
  {B{\"u}hler}, {Bulik}, {B{\"u}sching}, {Boutelier}, {Chadwick},
  {Charbonnier}, {Chaves}, {Cheesebrough}, {Chounet}, {Clapson}, {Coignet},
  {Dalton}, {Daniel}, {Davids}, {Degrange}, {Deil}, {Dickinson},
  {Djannati-Ata{\"i}}, {Domainko}, {Drury}, {Dubois}, {Dubus}, {Dyks}, {Dyrda},
  {Egberts}, {Emmanoulopoulos}, {Espigat}, {Farnier}, {Feinstein}, {Fiasson},
  {F{\"o}rster}, {Fontaine}, {F{\"u}{\ss}ling}, {Gabici}, {Gallant},
  {G{\'e}rard}, {Giebels}, {Glicenstein}, {Gl{\"u}ck}, {Goret}, {G{\"o}hring},
  {Hauser}, {Hauser}, {Heinz}, {Heinzelmann}, {Henri}, {Hermann}, {Hinton},
  {Hoffmann}, {Hofmann}, {Holleran}, {Hoppe}, {Horns}, {Jacholkowska}, {de
  Jager}, {Jahn}, {Jung}, {Katarzy{\'n}ski}, {Katz}, {Kaufmann}, {Kendziorra},
  {Kerschhaggl}, {Khangulyan}, {Kh{\'e}lifi}, {Keogh}, {Klu{\'z}niak},
  {Kneiske}, {Komin}, {Kosack}, {Lamanna}, {Latham}, {Lenain}, {Lohse},
  {Marandon}, {Martin}, {Martineau-Huynh}, {Marcowith}, {Maurin}, {McComb},
  {Medina}, {Moderski}, {Moulin}, {Naumann-Godo}, {de Naurois}, {Nedbal},
  {Nekrassov}, {Niemiec}, {Nolan}, {Ohm}, {Olive}, {de O{\~n}a Wilhelmi},
  {Orford}, {Ostrowski}, {Panter}, {Arribas}, {Pedaletti}, {Pelletier},
  {Petrucci}, {Pita}, {P{\"u}hlhofer}, {Punch}, {Quirrenbach}, {Raubenheimer},
  {Raue}, {Rayner}, {Renaud}, {Rieger}, {Ripken}, {Rob}, {Rosier-Lees},
  {Rowell}, {Rudak}, {Rulten}, {Ruppel}, {Sahakian}, {Santangelo},
  {Schlickeiser}, {Sch{\"o}ck}, {Schr{\"o}der}, {Schwanke}, {Schwarzburg},
  {Schwemmer}, {Shalchi}, {Sikora}, {Skilton}, {Sol}, {Spangler}, {Stawarz},
  {Steenkamp}, {Stegmann}, {Superina}, {Szostek}, {Tam}, {Tavernet}, {Terrier},
  {Tibolla}, {Tluczykont}, {van Eldik}, {Vasileiadis}, {Venter}, {Venter},
  {Vialle}, {Vincent}, {Vink}, {Vivier}, {V{\"o}lk}, {Volpe}, {Wagner}, {Ward},
  {Zdziarski}, \& {Zech}}]{2009ApJ...695L..40A}
{Aharonian}, F., {Akhperjanian}, A.~G., {Anton}, G., {et~al.} (\hess Collaboration)
  2009{\natexlab{b}}, \apjl, 695, L40

\bibitem[{{Aharonian} {et~al.}(2005{\natexlab{a}}){Aharonian}, {Akhperjanian},
  {Aye}, {Bazer-Bachi}, {Beilicke}, {Benbow}, {Berge}, {Berghaus},
  {Bernl{\"o}hr}, {Boisson}, {Bolz}, {Braun}, {Breitling}, {Brown}, {Bussons
  Gordo}, {Chadwick}, {Chounet}, {Cornils}, {Costamante}, {Degrange},
  {Djannati-Ata{\"i}}, {O'C.~Drury}, {Dubus}, {Emmanoulopoulos}, {Espigat},
  {Feinstein}, {Fleury}, {Fontaine}, {Fuchs}, {Funk}, {Gallant}, {Giebels},
  {Gillessen}, {Glicenstein}, {Goret}, {Hadjichristidis}, {Hauser},
  {Heinzelmann}, {Henri}, {Hermann}, {Hinton}, {Hofmann}, {Holleran}, {Horns},
  {de Jager}, {Kh{\'e}lifi}, {Komin}, {Konopelko}, {Latham}, {Le Gallou},
  {Lemi{\`e}re}, {Lemoine}, {Leroy}, {Lohse}, {Marcowith}, {Masterson},
  {McComb}, {de Naurois}, {Nolan}, {Noutsos}, {Orford}, {Osborne}, {Ouchrif},
  {Panter}, {Pelletier}, {Pita}, {P{\"u}hlhofer}, {Punch}, {Raubenheimer},
  {Raue}, {Raux}, {Rayner}, {Redondo}, {Reimer}, {Reimer}, {Ripken}, {Rob},
  {Rolland}, {Rowell}, {Sahakian}, {Saug{\'e}}, {Schlenker}, {Schlickeiser},
  {Schuster}, {Schwanke}, {Siewert}, {Sol}, {Steenkamp}, {Stegmann},
  {Tavernet}, {Terrier}, {Th{\'e}oret}, {Tluczykont}, {Vasileiadis}, {Venter},
  {Vincent}, {V{\"o}lk}, \& {Wagner}}]{2005A+A...437...95A}
{Aharonian}, F., {Akhperjanian}, A.~G., {Aye}, K.-M., {et~al.} (\hess Collaboration) 2005{\natexlab{a}}, \aap, 437, 95

\bibitem[{{Aharonian} {et~al.}(2005{\natexlab{b}}){Aharonian}, {Akhperjanian},
  {Aye}, {Bazer-Bachi}, {Beilicke}, {Benbow}, {Berge}, {Berghaus},
  {Bernl{\"o}hr}, {Bolz}, {Boisson}, {Borgmeier}, {Breitling}, {Brown},
  {Bussons Gordo}, {Chadwick}, {Chitnis}, {Chounet}, {Cornils}, {Costamante},
  {Degrange}, {Djannati-Ata{\"i}}, {Drury}, {Ergin}, {Espigat}, {Feinstein},
  {Fleury}, {Fontaine}, {Funk}, {Gallant}, {Giebels}, {Gillessen}, {Goret},
  {Guy}, {Hadjichristidis}, {Hauser}, {Heinzelmann}, {Henri}, {Hermann},
  {Hinton}, {Hofmann}, {Holleran}, {Horns}, {de Jager}, {Jung I.},
  {Kh{\'e}lifi}, {Komin}, {Konopelko}, {Latham}, {Le Gallou}, {Lemoine},
  {Lemi{\`e}re}, {Leroy}, {Lohse}, {Marcowith}, {Masterson}, {McComb}, {de
  Naurois}, {Nolan}, {Noutsos}, {Orford}, {Osborne}, {Ouchrif}, {Panter},
  {Pelletier}, {Pita}, {Pohl}, {P{\"u}hlhofer}, {Punch}, {Raubenheimer},
  {Raue}, {Raux}, {Rayner}, {Redondo}, {Reimer}, {Reimer}, {Ripken}, {Rivoal},
  {Rob}, {Rolland}, {Rowell}, {Sahakian}, {Saug{\'e}}, {Schlenker},
  {Schlickeiser}, {Schuster}, {Schwanke}, {Siewert}, {Sol}, {Steenkamp},
  {Stegmann}, {Tavernet}, {Th{\'e}oret}, {Tluczykont}, {van der Walt},
  {Vasileiadis}, {Vincent}, {Visser}, {V{\"o}lk}, \&
  {Wagner}}]{2005A+A...430..865A}
{Aharonian}, F., {Akhperjanian}, A.~G., {Aye}, K.-M., {et~al.} (\hess Collaboration)
  2005{\natexlab{b}}, \aap, 430, 865

\bibitem[{{Aharonian} {et~al.}(2006{\natexlab{a}}){Aharonian}, {Akhperjanian},
  {Bazer-Bachi}, {Beilicke}, {Benbow}, {Berge}, {Bernl{\"o}hr}, {Boisson},
  {Bolz}, {Borrel}, {Braun}, {Breitling}, {Brown}, {B{\"u}hler},
  {B{\"u}sching}, {Carrigan}, {Chadwick}, {Chounet}, {Cornils}, {Costamante},
  {Degrange}, {Dickinson}, {Djannati-Ata{\"i}}, {O'C.~Drury}, {Dubus},
  {Egberts}, {Emmanoulopoulos}, {Espigat}, {Feinstein}, {Ferrero}, {Fiasson},
  {Fontaine}, {Funk}, {Funk}, {Gallant}, {Giebels}, {Glicenstein}, {Goret},
  {Hadjichristidis}, {Hauser}, {Hauser}, {Heinzelmann}, {Henri}, {Hermann},
  {Hinton}, {Hofmann}, {Holleran}, {Horns}, {Jacholkowska}, {de Jager},
  {Kh{\'e}lifi}, {Komin}, {Konopelko}, {Kosack}, {Latham}, {Le Gallou},
  {Lemi{\`e}re}, {Lemoine-Goumard}, {Lohse}, {Martin}, {Martineau-Huynh},
  {Marcowith}, {Masterson}, {McComb}, {de Naurois}, {Nedbal}, {Nolan},
  {Noutsos}, {Orford}, {Osborne}, {Ouchrif}, {Panter}, {Pelletier}, {Pita},
  {P{\"u}hlhofer}, {Punch}, {Raubenheimer}, {Raue}, {Rayner}, {Reimer},
  {Reimer}, {Ripken}, {Rob}, {Rolland}, {Rowell}, {Sahakian}, {Saug{\'e}},
  {Schlenker}, {Schlickeiser}, {Schwanke}, {Sol}, {Spangler}, {Spanier},
  {Steenkamp}, {Stegmann}, {Superina}, {Tavernet}, {Terrier}, {Th{\'e}oret},
  {Tluczykont}, {van Eldik}, {Vasileiadis}, {Venter}, {Vincent}, {V{\"o}lk},
  {Wagner}, \& {Ward}}]{2006A+A...457..899A}
{Aharonian}, F., {Akhperjanian}, A.~G., {Bazer-Bachi}, A.~R., {et~al.} (\hess Collaboration)
  2006{\natexlab{a}}, \aap, 457, 899

\bibitem[{{Aharonian} {et~al.}(2006{\natexlab{b}}){Aharonian}, {Akhperjanian},
  {Bazer-Bachi}, {Beilicke}, {Benbow}, {Berge}, {Bernl{\"o}hr}, {Boisson},
  {Bolz}, {Borrel}, {Braun}, {Breitling}, {Brown}, {Chadwick}, {Chounet},
  {Cornils}, {Costamante}, {Degrange}, {Dickinson}, {Djannati-Ata{\"i}},
  {Drury}, {Dubus}, {Emmanoulopoulos}, {Espigat}, {Feinstein}, {Fontaine},
  {Fuchs}, {Funk}, {Gallant}, {Giebels}, {Gillessen}, {Glicenstein}, {Goret},
  {Hadjichristidis}, {Hauser}, {Hauser}, {Heinzelmann}, {Henri}, {Hermann},
  {Hinton}, {Hofmann}, {Holleran}, {Horns}, {Jacholkowska}, {de Jager},
  {Kh{\'e}lifi}, {Klages}, {Komin}, {Konopelko}, {Latham}, {Le Gallou},
  {Lemi{\`e}re}, {Lemoine-Goumard}, {Leroy}, {Lohse}, {Martin},
  {Martineau-Huynh}, {Marcowith}, {Masterson}, {McComb}, {de Naurois}, {Nolan},
  {Noutsos}, {Orford}, {Osborne}, {Ouchrif}, {Panter}, {Pelletier}, {Pita},
  {P{\"u}hlhofer}, {Punch}, {Raubenheimer}, {Raue}, {Raux}, {Rayner}, {Reimer},
  {Reimer}, {Ripken}, {Rob}, {Rolland}, {Rowell}, {Sahakian}, {Saug{\'e}},
  {Schlenker}, {Schlickeiser}, {Schuster}, {Schwanke}, {Siewert}, {Sol},
  {Spangler}, {Steenkamp}, {Stegmann}, {Tavernet}, {Terrier}, {Th{\'e}oret},
  {Tluczykont}, {van Eldik}, {Vasileiadis}, {Venter}, {Vincent}, {V{\"o}lk}, \&
  {Wagner}}]{2006Natur.440.1018A}
{Aharonian}, F., {Akhperjanian}, A.~G., {Bazer-Bachi}, A.~R., {et~al.} (\hess Collaboration)
  2006{\natexlab{b}}, \nat, 440, 1018

\bibitem[{{Aharonian}(2000)}]{2000NewA....5..377A}
{Aharonian}, F.~A. 2000, New Astronomy, 5, 377

\bibitem[{{Albert} {et~al.}(2007){Albert}, {Aliu}, {Anderhub}, {Antoranz},
  {Armada}, {Baixeras}, {Barrio}, {Bartko}, {Bastieri}, {Becker}, {Bednarek},
  {Berger}, {Bigongiari}, {Biland}, {Bock}, {Bordas}, {Bosch-Ramon}, {Bretz},
  {Britvitch}, {Camara}, {Carmona}, {Chilingarian}, {Coarasa}, {Commichau},
  {Contreras}, {Cortina}, {Costado}, {Curtef}, {Danielyan}, {Dazzi}, {De
  Angelis}, {Delgado}, {de los Reyes}, {De Lotto}, {Domingo-Santamar{\'{\i}}a},
  {Dorner}, {Doro}, {Errando}, {Fagiolini}, {Ferenc}, {Fern{\'a}ndez}, {Firpo},
  {Flix}, {Fonseca}, {Font}, {Fuchs}, {Galante}, {Garc{\'{\i}}a-L{\'o}pez},
  {Garczarczyk}, {Gaug}, {Giller}, {Goebel}, {Hakobyan}, {Hayashida},
  {Hengstebeck}, {Herrero}, {H{\"o}hne}, {Hose}, {Hsu}, {Jacon}, {Jogler},
  {Kosyra}, {Kranich}, {Kritzer}, {Laille}, {Lindfors}, {Lombardi}, {Longo},
  {L{\'o}pez}, {L{\'o}pez}, {Lorenz}, {Majumdar}, {Maneva}, {Mannheim},
  {Mansutti}, {Mariotti}, {Mart{\'{\i}}nez}, {Mazin}, {Merck}, {Meucci},
  {Meyer}, {Miranda}, {Mirzoyan}, {Mizobuchi}, {Moralejo}, {Nieto}, {Nilsson},
  {Ninkovic}, {O{\~n}a-Wilhelmi}, {Otte}, {Oya}, {Paneque}, {Panniello},
  {Paoletti}, {Paredes}, {Pasanen}, {Pascoli}, {Pauss}, {Pegna}, {Perlman},
  {Persic}, {Peruzzo}, {Piccioli}, {Prandini}, {Puchades}, {Raymers}, {Rhode},
  {Rib{\'o}}, {Rico}, {Rissi}, {Robert}, {R{\"u}gamer}, {Saggion}, {Saito},
  {S{\'a}nchez}, {Sartori}, {Scalzotto}, {Scapin}, {Schmitt}, {Schweizer},
  {Shayduk}, {Shinozaki}, {Shore}, {Sidro}, {Sillanp{\"a}{\"a}}, {Sobczynska},
  {Stamerra}, {Stark}, {Takalo}, {Tavecchio}, {Temnikov}, {Tescaro}, {Teshima},
  {Torres}, {Turini}, {Vankov}, {Vitale}, {Wagner}, {Wibig}, {Wittek},
  {Zandanel}, {Zanin}, \& {Zapatero}}]{2007ApJ...667L..21A}
{Albert}, J., {Aliu}, E., {Anderhub}, H., {et~al.} (MAGIC Collaboration) 2007, \apjl, 667, L21

\bibitem[{{Aleksi{\'c}} {et~al.}(2012){Aleksi{\'c}}, {Alvarez}, {Antonelli},
  {Antoranz}, {Asensio}, {Backes}, {Barres de Almeida}, {Barrio}, {Bastieri},
  {Becerra Gonz{\'a}lez}, {Bednarek}, {Berger}, {Bernardini}, {Biland},
  {Blanch}, {Bock}, {Boller}, {Bonnoli}, {Borla Tridon}, {Bretz},
  {Ca{\~n}ellas}, {Carmona}, {Carosi}, {Colin}, {Colombo}, {Contreras},
  {Cortina}, {Cossio}, {Covino}, {da Vela}, {Dazzi}, {de Angelis}, {de Caneva},
  {de Cea Del Pozo}, {de Lotto}, {Delgado Mendez}, {Diago Ortega}, {Doert},
  {Dom{\'{\i}}nguez}, {Dominis Prester}, {Dorner}, {Doro}, {Eisenacher},
  {Elsaesser}, {Ferenc}, {Fonseca}, {Font}, {Fruck}, {Garc{\'{\i}}a L{\'o}pez},
  {Garczarczyk}, {Garrido}, {Giavitto}, {Godinovi{\'c}}, {Gozzini}, {Hadasch},
  {H{\"a}fner}, {Herrero}, {Hildebrand}, {H{\"o}hne-M{\"o}nch}, {Hose},
  {Hrupec}, {Huber}, {Jogler}, {Kadenius}, {Kellermann}, {Klepser},
  {Kr{\"a}henb{\"u}hl}, {Krause}, {La Barbera}, {Lelas}, {Leonardo},
  {Lewandowska}, {Lindfors}, {Lombardi}, {L{\'o}pez}, {L{\'o}pez-Coto},
  {L{\'o}pez-Oramas}, {Lorenz}, {Makariev}, {Maneva}, {Mankuzhiyil},
  {Mannheim}, {Maraschi}, {Mariotti}, {Mart{\'{\i}}nez}, {Mazin}, {Meucci},
  {Miranda}, {Mirzoyan}, {Mold{\'o}n}, {Moralejo}, {Munar-Adrover},
  {Niedzwiecki}, {Nieto}, {Nilsson}, {Nowak}, {Orito}, {Paiano}, {Paneque},
  {Paoletti}, {Pardo}, {Paredes}, {Partini}, {Perez-Torres}, {Persic},
  {Peruzzo}, {Pilia}, {Pochon}, {Prada}, {Prada Moroni}, {Prandini}, {Puerto
  Gimenez}, {Puljak}, {Reichardt}, {Reinthal}, {Rhode}, {Rib{\'o}}, {Rico},
  {R{\"u}gamer}, {Saggion}, {Saito}, {Saito}, {Salvati}, {Satalecka},
  {Scalzotto}, {Scapin}, {Schultz}, {Schweizer}, {Shayduk}, {Shore},
  {Sillanp{\"a}{\"a}}, {Sitarek}, {Snidaric}, {Sobczynska}, {Spanier}, {Spiro},
  {Stamatescu}, {Stamerra}, {Steinke}, {Storz}, {Strah}, {Sun}, {Suri{\'c}},
  {Takalo}, {Takami}, {Tavecchio}, {Temnikov}, {Terzi{\'c}}, {Tescaro},
  {Teshima}, {Tibolla}, {Torres}, {Treves}, {Uellenbeck}, {Vogler}, {Wagner},
  {Weitzel}, {Zabalza}, {Zandanel}, {Zanin}, {Pfrommer}, \&
  {Pinzke}}]{2012A+A...539L...2A}
{Aleksi{\'c}}, J., {Alvarez}, E.~A., {Antonelli}, L.~A., {et~al.} (MAGIC Collaboration) 2012, \aap,
  539, L2

\bibitem[{{Aleksi{\'c}} {et~al.}(2011){Aleksi{\'c}}, {Antonelli}, {Antoranz},
  {Backes}, {Barrio}, {Bastieri}, {Becerra Gonz{\'a}lez}, {Bednarek},
  {Berdyugin}, {Berger}, {Bernardini}, {Biland}, {Blanch}, {Bock}, {Boller},
  {Bonnoli}, {Borla Tridon}, {Braun}, {Bretz}, {Ca{\~n}ellas}, {Carmona},
  {Carosi}, {Colin}, {Colombo}, {Contreras}, {Cortina}, {Cossio}, {Covino},
  {Dazzi}, {De Angelis}, {De Cea del Pozo}, {De Lotto}, {Delgado Mendez},
  {Diago Ortega}, {Doert}, {Dom{\'{\i}}nguez}, {Dominis Prester}, {Dorner},
  {Doro}, {Elsaesser}, {Ferenc}, {Fonseca}, {Font}, {Fruck}, {Garc{\'{\i}}a
  L{\'o}pez}, {Garczarczyk}, {Garrido}, {Giavitto}, {Godinovi{\'c}}, {Hadasch},
  {H{\"a}fner}, {Herrero}, {Hildebrand}, {H{\"o}hne-M{\"o}nch}, {Hose},
  {Hrupec}, {Huber}, {Jogler}, {Klepser}, {Kr{\"a}henb{\"u}hl}, {Krause}, {La
  Barbera}, {Lelas}, {Leonardo}, {Lindfors}, {Lombardi}, {L{\'o}pez}, {Lorenz},
  {Makariev}, {Maneva}, {Mankuzhiyil}, {Mannheim}, {Maraschi}, {Mariotti},
  {Mart{\'{\i}}nez}, {Mazin}, {Meucci}, {Miranda}, {Mirzoyan}, {Miyamoto},
  {Mold{\'o}n}, {Moralejo}, {Nieto}, {Nilsson}, {Orito}, {Oya}, {Paneque},
  {Paoletti}, {Pardo}, {Paredes}, {Partini}, {Pasanen}, {Pauss},
  {Perez-Torres}, {Persic}, {Peruzzo}, {Pilia}, {Pochon}, {Prada}, {Prada
  Moroni}, {Prandini}, {Puljak}, {Reichardt}, {Reinthal}, {Rhode}, {Rib{\'o}},
  {Rico}, {R{\"u}gamer}, {Saggion}, {Saito}, {Saito}, {Salvati}, {Satalecka},
  {Scalzotto}, {Scapin}, {Schultz}, {Schweizer}, {Shayduk}, {Shore},
  {Sillanp{\"a}{\"a}}, {Sitarek}, {Sobczynska}, {Spanier}, {Spiro}, {Stamerra},
  {Steinke}, {Storz}, {Strah}, {Suri{\'c}}, {Takalo}, {Tavecchio}, {Temnikov},
  {Terzi{\'c}}, {Tescaro}, {Teshima}, {Thom}, {Tibolla}, {Torres}, {Treves},
  {Vankov}, {Vogler}, {Wagner}, {Weitzel}, {Zabalza}, {Zandanel}, {Zanin},
  {MAGIC Collaboration}, {Tanaka}, {Wood}, \& {Buson}}]{2011ApJ...730L...8A}
{Aleksi{\'c}}, J., {Antonelli}, L.~A., {Antoranz}, P., {et~al.} (MAGIC Collaboration) 2011, \apjl,
  730, L8+

\bibitem[{{Allen} {et~al.}(1982){Allen}, {Ward}, \&
  {Hyland}}]{1982MNRAS.199..969A}
{Allen}, D.~A., {Ward}, M.~J., \& {Hyland}, A.~R. 1982, \mnras, 199, 969

\bibitem[{{Atwood} {et~al.}(2009){Atwood}, {Abdo}, {Ackermann}, {Althouse},
  {Anderson}, {Axelsson}, {Baldini}, {Ballet}, {Band}, {Barbiellini},
  {Bartelt}, {Bastieri}, {Baughman}, {Bechtol}, {B{\'e}d{\'e}r{\`e}de},
  {Bellardi}, {Bellazzini}, {Berenji}, {Bignami}, {Bisello}, {Bissaldi},
  {Blandford}, {Bloom}, {Bogart}, {Bonamente}, {Bonnell}, {Borgland},
  {Bouvier}, {Bregeon}, {Brez}, {Brigida}, {Bruel}, {Burnett}, {Busetto},
  {Caliandro}, {Cameron}, {Caraveo}, {Carius}, {Carlson}, {Casandjian},
  {Cavazzuti}, {Ceccanti}, {Cecchi}, {Charles}, {Chekhtman}, {Cheung},
  {Chiang}, {Chipaux}, {Cillis}, {Ciprini}, {Claus}, {Cohen-Tanugi},
  {Condamoor}, {Conrad}, {Corbet}, {Corucci}, {Costamante}, {Cutini}, {Davis},
  {Decotigny}, {DeKlotz}, {Dermer}, {de Angelis}, {Digel}, {do Couto e Silva},
  {Drell}, {Dubois}, {Dumora}, {Edmonds}, {Fabiani}, {Farnier}, {Favuzzi},
  {Flath}, {Fleury}, {Focke}, {Funk}, {Fusco}, {Gargano}, {Gasparrini},
  {Gehrels}, {Gentit}, {Germani}, {Giebels}, {Giglietto}, {Giommi}, {Giordano},
  {Glanzman}, {Godfrey}, {Grenier}, {Grondin}, {Grove}, {Guillemot}, {Guiriec},
  {Haller}, {Harding}, {Hart}, {Hays}, {Healey}, {Hirayama}, {Hjalmarsdotter},
  {Horn}, {Hughes}, {J{\'o}hannesson}, {Johansson}, {Johnson}, {Johnson},
  {Johnson}, {Johnson}, {Kamae}, {Katagiri}, {Kataoka}, {Kavelaars}, {Kawai},
  {Kelly}, {Kerr}, {Klamra}, {Kn{\"o}dlseder}, {Kocian}, {Komin}, {Kuehn},
  {Kuss}, {Landriu}, {Latronico}, {Lee}, {Lee}, {Lemoine-Goumard}, {Lionetto},
  {Longo}, {Loparco}, {Lott}, {Lovellette}, {Lubrano}, {Madejski}, {Makeev},
  {Marangelli}, {Massai}, {Mazziotta}, {McEnery}, {Menon}, {Meurer},
  {Michelson}, {Minuti}, {Mirizzi}, {Mitthumsiri}, {Mizuno}, {Moiseev},
  {Monte}, {Monzani}, {Moretti}, {Morselli}, {Moskalenko}, {Murgia},
  {Nakamori}, {Nishino}, {Nolan}, {Norris}, {Nuss}, {Ohno}, {Ohsugi}, {Omodei},
  {Orlando}, {Ormes}, {Paccagnella}, {Paneque}, {Panetta}, {Parent}, {Pearce},
  {Pepe}, {Perazzo}, {Pesce-Rollins}, {Picozza}, {Pieri}, {Pinchera}, {Piron},
  {Porter}, {Poupard}, {Rain{\`o}}, {Rando}, {Rapposelli}, {Razzano}, {Reimer},
  {Reimer}, {Reposeur}, {Reyes}, {Ritz}, {Rochester}, {Rodriguez}, {Romani},
  {Roth}, {Russell}, {Ryde}, {Sabatini}, {Sadrozinski}, {Sanchez}, {Sander},
  {Sapozhnikov}, {Parkinson}, {Scargle}, {Schalk}, {Scolieri}, {Sgr{\`o}},
  {Share}, {Shaw}, {Shimokawabe}, {Shrader}, {Sierpowska-Bartosik}, {Siskind},
  {Smith}, {Smith}, {Spandre}, {Spinelli}, {Starck}, {Stephens}, {Strickman},
  {Strong}, {Suson}, {Tajima}, {Takahashi}, {Takahashi}, {Tanaka}, {Tenze},
  {Tether}, {Thayer}, {Thayer}, {Thompson}, {Tibaldo}, {Tibolla}, {Torres},
  {Tosti}, {Tramacere}, {Turri}, {Usher}, {Vilchez}, {Vitale}, {Wang},
  {Watters}, {Winer}, {Wood}, {Ylinen}, \& {Ziegler}}]{2009ApJ...697.1071A}
{Atwood}, W.~B., {Abdo}, A.~A., {Ackermann}, M., {et~al.} (\fermi Collaboration) 2009, \apj, 697, 1071

\bibitem[{{Becherini} {et~al.}(2011){Becherini}, {Djannati-Ata{\"i}},
  {Marandon}, {Punch}, \& {Pita}}]{2011APh....34..858B}
{Becherini}, Y., {Djannati-Ata{\"i}}, A., {Marandon}, V., {Punch}, M., \&
  {Pita}, S. 2011, Astroparticle Physics, 34, 858

\bibitem[{{Cannon}(2010)}]{2010ATel.2591....1C}
{Cannon}, A. (for the \fermi Collaboration) 2010, The Astronomer's Telegram, 2591

\bibitem[{{Cerruti} {et~al.}(2013){Cerruti}, {Dermer}, {Lott}, {Boisson}, \&
  {Zech}}]{2013ApJ...771L...4C}
{Cerruti}, M., {Dermer}, C.~D., {Lott}, B., {Boisson}, C., \& {Zech}, A. 2013,
  \apjl, 771, L4

\bibitem[{{Chen} {et~al.}(2005){Chen}, {Fu}, \& {Gao}}]{2005NewA...11...27C}
{Chen}, P.~S., {Fu}, H.~W., \& {Gao}, Y.~F. 2005, \na, 11, 27

\bibitem[{{Cohen} {et~al.}(2007){Cohen}, {Lane}, {Cotton}, {Kassim}, {Lazio},
  {Perley}, {Condon}, \& {Erickson}}]{2007AJ....134.1245C}
{Cohen}, A.~S., {Lane}, W.~M., {Cotton}, W.~D., {et~al.} 2007, \aj, 134, 1245

\bibitem[{{Cutri} {et~al.}(2003){Cutri}, {Skrutskie}, {van Dyk}, {Beichman},
  {Carpenter}, {Chester}, {Cambresy}, {Evans}, {Fowler}, {Gizis}, {Howard},
  {Huchra}, {Jarrett}, {Kopan}, {Kirkpatrick}, {Light}, {Marsh}, {McCallon},
  {Schneider}, {Stiening}, {Sykes}, {Weinberg}, {Wheaton}, {Wheelock}, \&
  {Zacarias}}]{2003yCat.2246....0C}
{Cutri}, R.~M., {Skrutskie}, M.~F., {van Dyk}, S., {et~al.} 2003, VizieR Online
  Data Catalog, 2246, 0

\bibitem[{{de Naurois} \& {Rolland}(2009)}]{2009APh....32..231D}
{de Naurois}, M. \& {Rolland}, L. 2009, Astroparticle Physics, 32, 231

\bibitem[{{Dole} {et~al.}(2006){Dole}, {Lagache}, {Puget}, {Caputi},
  {Fern{\'a}ndez-Conde}, {Le Floc'h}, {Papovich}, {P{\'e}rez-Gonz{\'a}lez},
  {Rieke}, \& {Blaylock}}]{2006A+A...451..417D}
{Dole}, H., {Lagache}, G., {Puget}, J.-L., {et~al.} 2006, \aap, 451, 417

\bibitem[{{Dom{\'{\i}}nguez} {et~al.}(2011){Dom{\'{\i}}nguez}, {Primack},
  {Rosario}, {Prada}, {Gilmore}, {Faber}, {Koo}, {Somerville},
  {P{\'e}rez-Torres}, {P{\'e}rez-Gonz{\'a}lez}, {Huang}, {Davis},
  {Guhathakurta}, {Barmby}, {Conselice}, {Lozano}, {Newman}, \&
  {Cooper}}]{2011MNRAS.410.2556D}
{Dom{\'{\i}}nguez}, A., {Primack}, J.~R., {Rosario}, D.~J., {et~al.} 2011,
  \mnras, 410, 2556

\bibitem[{{Douglas} {et~al.}(1996){Douglas}, {Bash}, {Bozyan}, {Torrence}, \&
  {Wolfe}}]{1996AJ....111.1945D}
{Douglas}, J.~N., {Bash}, F.~N., {Bozyan}, F.~A., {Torrence}, G.~W., \&
  {Wolfe}, C. 1996, \aj, 111, 1945

\bibitem[{{Dwek} \& {Krennrich}(2013)}]{2013APh....43..112D}
{Dwek}, E. \& {Krennrich}, F. 2013, Astroparticle Physics, 43, 112

\bibitem[{{Edelson}(1992)}]{1992ApJ...401..516E}
{Edelson}, R. 1992, \apj, 401, 516

\bibitem[{{Falomo} \& {Ulrich}(2000)}]{2000A+A...357...91F}
{Falomo}, R. \& {Ulrich}, M.-H. 2000, \aap, 357, 91

\bibitem[{{Fazio} \& {Stecker}(1970)}]{1970Natur.226..135F}
{Fazio}, G.~G. \& {Stecker}, F.~W. 1970, \nat, 226, 135

\bibitem[{{Feldman} \& {Cousins}(1998)}]{1998PhRvD..57.3873F}
{Feldman}, G.~J. \& {Cousins}, R.~D. 1998, \prd, 57, 3873

\bibitem[{{Finke} {et~al.}(2008){Finke}, {Dermer}, \&
  {B{\"o}ttcher}}]{2008ApJ...686..181F}
{Finke}, J.~D., {Dermer}, C.~D., \& {B{\"o}ttcher}, M. 2008, \apj, 686, 181

\bibitem[{{Fossati} {et~al.}(2000{\natexlab{a}}){Fossati}, {Celotti},
  {Chiaberge}, {Zhang}, {Chiappetti}, {Ghisellini}, {Maraschi}, {Tavecchio},
  {Pian}, \& {Treves}}]{2000ApJ...541..153F}
{Fossati}, G., {Celotti}, A., {Chiaberge}, M., {et~al.} 2000{\natexlab{a}},
  \apj, 541, 153

\bibitem[{{Fossati} {et~al.}(2000{\natexlab{b}}){Fossati}, {Celotti},
  {Chiaberge}, {Zhang}, {Chiappetti}, {Ghisellini}, {Maraschi}, {Tavecchio},
  {Pian}, \& {Treves}}]{2000ApJ...541..166F}
{Fossati}, G., {Celotti}, A., {Chiaberge}, M., {et~al.} 2000{\natexlab{b}},
  \apj, 541, 166

\bibitem[{{Franceschini} {et~al.}(2008){Franceschini}, {Rodighiero}, \&
  {Vaccari}}]{2008A+A...487..837F}
{Franceschini}, A., {Rodighiero}, G., \& {Vaccari}, M. 2008, \aap, 487, 837

\bibitem[{{Gehrels} {et~al.}(2004){Gehrels}, {Chincarini}, {Giommi}, {Mason},
  {Nousek}, {Wells}, {White}, {Barthelmy}, {Burrows}, {Cominsky}, {Hurley},
  {Marshall}, {M{\'e}sz{\'a}ros}, {Roming}, {Angelini}, {Barbier}, {Belloni},
  {Campana}, {Caraveo}, {Chester}, {Citterio}, {Cline}, {Cropper}, {Cummings},
  {Dean}, {Feigelson}, {Fenimore}, {Frail}, {Fruchter}, {Garmire}, {Gendreau},
  {Ghisellini}, {Greiner}, {Hill}, {Hunsberger}, {Krimm}, {Kulkarni}, {Kumar},
  {Lebrun}, {Lloyd-Ronning}, {Markwardt}, {Mattson}, {Mushotzky}, {Norris},
  {Osborne}, {Paczynski}, {Palmer}, {Park}, {Parsons}, {Paul}, {Rees},
  {Reynolds}, {Rhoads}, {Sasseen}, {Schaefer}, {Short}, {Smale}, {Smith},
  {Stella}, {Tagliaferri}, {Takahashi}, {Tashiro}, {Townsley}, {Tueller},
  {Turner}, {Vietri}, {Voges}, {Ward}, {Willingale}, {Zerbi}, \&
  {Zhang}}]{2004ApJ...611.1005G}
{Gehrels}, N., {Chincarini}, G., {Giommi}, P., {et~al.} 2004, \apj, 611, 1005

\bibitem[{{Georganopoulos} {et~al.}(2010){Georganopoulos}, {Finke}, \&
  {Reyes}}]{2010ApJ...714L.157G}
{Georganopoulos}, M., {Finke}, J.~D., \& {Reyes}, L.~C. 2010, \apjl, 714, L157

\bibitem[{{Ghisellini} \& {Maraschi}(1989)}]{1989ApJ...340..181G}
{Ghisellini}, G. \& {Maraschi}, L. 1989, \apj, 340, 181

\bibitem[{{Ghisellini} {et~al.}(1996){Ghisellini}, {Maraschi}, \&
  {Dondi}}]{1996A+AS..120C.503G}
{Ghisellini}, G., {Maraschi}, L., \& {Dondi}, L. 1996, \aaps, 120, C503+

\bibitem[{{Ginzburg} \& {Syrovatskii}(1965)}]{1965ARA+A...3..297G}
{Ginzburg}, V.~L. \& {Syrovatskii}, S.~I. 1965, \araa, 3, 297

\bibitem[{{Gould} \& {Schr{\'e}der}(1967)}]{1967PhRv..155.1408G}
{Gould}, R.~J. \& {Schr{\'e}der}, G.~P. 1967, Physical Review, 155, 1408

\bibitem[{{Griffith} {et~al.}(1994){Griffith}, {Wright}, {Burke}, \&
  {Ekers}}]{1994ApJS...90..179G}
{Griffith}, M.~R., {Wright}, A.~E., {Burke}, B.~F., \& {Ekers}, R.~D. 1994,
  \apjs, 90, 179

\bibitem[{{Hauser} {et~al.}(2004){Hauser}, {M{\"o}llenhoff}, {P{\"u}hlhofer},
  {Wagner}, {Hagen}, \& {Knoll}}]{2004AN....325..659H}
{Hauser}, M., {M{\"o}llenhoff}, C., {P{\"u}hlhofer}, G., {et~al.} 2004,
  Astronomische Nachrichten, 325, 659

\bibitem[{{Hauser} {et~al.}(1998){Hauser}, {Arendt}, {Kelsall}, {Dwek},
  {Odegard}, {Weiland}, {Freudenreich}, {Reach}, {Silverberg}, {Moseley},
  {Pei}, {Lubin}, {Mather}, {Shafer}, {Smoot}, {Weiss}, {Wilkinson}, \&
  {Wright}}]{1998ApJ...508...25H}
{Hauser}, M.~G., {Arendt}, R.~G., {Kelsall}, T., {et~al.} 1998, \apj, 508, 25

\bibitem[{{Hauser} \& {Dwek}(2001)}]{2001ARA+A..39..249H}
{Hauser}, M.~G. \& {Dwek}, E. 2001, \araa, 39, 249

\bibitem[{{H.E.S.S.~Collaboration}
  {et~al.}(2013{\natexlab{a}}){H.E.S.S.~Collaboration}, {Abramowski}, {Acero},
  {Aharonian}, {Akhperjanian}, {Anton}, {Balenderan}, {Balzer}, {Barnacka},
  {Becherini}, {Becker Tjus}, {Behera}, {Bernl{\"o}hr}, {Birsin}, {Biteau},
  {Bochow}, {Boisson}, {Bolmont}, {Bordas}, {Brucker}, {Brun}, {Brun}, {Bulik},
  {Carrigan}, {Casanova}, {Cerruti}, {Chadwick}, {Chaves}, {Cheesebrough},
  {Colafrancesco}, {Cologna}, {Conrad}, {Couturier}, {Dalton}, {Daniel},
  {Davids}, {Degrange}, {Deil}, {deWilt}, {Dickinson}, {Djannati-Ata{\"i}},
  {Domainko}, {O'C.~Drury}, {Dubus}, {Dutson}, {Dyks}, {Dyrda}, {Egberts},
  {Eger}, {Espigat}, {Fallon}, {Farnier}, {Fegan}, {Feinstein}, {Fernandes},
  {Fernandez}, {Fiasson}, {Fontaine}, {F{\"o}rster}, {F{\"u}{\ss}ling},
  {Gajdus}, {Gallant}, {Garrigoux}, {Gast}, {Giebels}, {Glicenstein},
  {Gl{\"u}ck}, {G{\"o}ring}, {Grondin}, {Grudzi{\'n}ska}, {H{\"a}ffner},
  {Hague}, {Hahn}, {Hampf}, {Harris}, {Hauser}, {Heinz}, {Heinzelmann},
  {Henri}, {Hermann}, {Hillert}, {Hinton}, {Hofmann}, {Hofverberg}, {Holler},
  {Horns}, {Jacholkowska}, {Jahn}, {Jamrozy}, {Jung}, {Kastendieck},
  {Katarzy{\'n}ski}, {Katz}, {Kaufmann}, {Kh{\'e}lifi}, {Klepser}, {Klochkov},
  {Klu{\'z}niak}, {Kneiske}, {Kolitzus}, {Komin}, {Kosack}, {Kossakowski},
  {Krayzel}, {Kr{\"u}ger}, {Laffon}, {Lamanna}, {Lefaucheur},
  {Lemoine-Goumard}, {Lenain}, {Lennarz}, {Lohse}, {Lopatin}, {Lu}, {Marandon},
  {Marcowith}, {Masbou}, {Maurin}, {Maxted}, {Mayer}, {McComb}, {Medina},
  {M{\'e}hault}, {Menzler}, {Moderski}, {Mohamed}, {Moulin}, {Naumann},
  {Naumann-Godo}, {de Naurois}, {Nedbal}, {Nguyen}, {Niemiec}, {Nolan}, {Ohm},
  {de O{\~n}a Wilhelmi}, {Opitz}, {Ostrowski}, {Oya}, {Panter}, {Parsons}, {Paz
  Arribas}, {Pekeur}, {Pelletier}, {Perez}, {Petrucci}, {Peyaud}, {Pita},
  {P{\"u}hlhofer}, {Punch}, {Quirrenbach}, {Raab}, {Raue}, {Reimer}, {Reimer},
  {Renaud}, {de los Reyes}, {Rieger}, {Ripken}, {Rob}, {Rosier-Lees}, {Rowell},
  {Rudak}, {Rulten}, {Sahakian}, {Sanchez}, {Santangelo}, {Schlickeiser},
  {Schulz}, {Schwanke}, {Schwarzburg}, {Schwemmer}, {Sheidaei}, {Skilton},
  {Sol}, {Spengler}, {Stawarz}, {Steenkamp}, {Stegmann}, {Stinzing}, {Stycz},
  {Sushch}, {Szostek}, {Tavernet}, {Terrier}, {Tluczykont}, {Trichard},
  {Valerius}, {van Eldik}, {Vasileiadis}, {Venter}, {Viana}, {Vincent},
  {V{\"o}lk}, {Volpe}, {Vorobiov}, {Vorster}, {Wagner}, {Ward}, {White},
  {Wierzcholska}, {Wouters}, {Zacharias}, {Zajczyk}, {Zdziarski}, {Zech}, \&
  {Zechlin}}]{2013A+A...554A.107H}
{H.E.S.S.~Collaboration}, {Abramowski}, A., {Acero}, F., {et~al.}
  2013{\natexlab{a}}, \aap, 554, A107

\bibitem[{{H.E.S.S.~Collaboration}
  {et~al.}(2013{\natexlab{b}}){H.E.S.S.~Collaboration}, {Abramowski}, {Acero},
  {Aharonian}, {Akhperjanian}, {Anton}, {Balenderan}, {Balzer}, {Barnacka},
  {Becherini}, {Becker Tjus}, {Bernl{\"o}hr}, {Birsin}, {Biteau}, {Bochow},
  {Boisson}, {Bolmont}, {Bordas}, {Brucker}, {Brun}, {Brun}, {Bulik},
  {Carrigan}, {Casanova}, {Cerruti}, {Chadwick}, {Charbonnier}, {Chaves},
  {Cheesebrough}, {Cologna}, {Conrad}, {Couturier}, {Dalton}, {Daniel},
  {Davids}, {Degrange}, {Deil}, {deWilt}, {Dickinson}, {Djannati-Ata{\"i}},
  {Domainko}, {O'C.~Drury}, {Dubus}, {Dutson}, {Dyks}, {Dyrda}, {Egberts},
  {Eger}, {Espigat}, {Fallon}, {Farnier}, {Fegan}, {Feinstein}, {Fernandes},
  {Fernandez}, {Fiasson}, {Fontaine}, {F{\"o}rster}, {F{\"u}{\ss}ling},
  {Gajdus}, {Gallant}, {Garrigoux}, {Gast}, {Giebels}, {Glicenstein},
  {Gl{\"u}ck}, {G{\"o}ring}, {Grondin}, {H{\"a}ffner}, {Hague}, {Hahn},
  {Hampf}, {Harris}, {Heinz}, {Heinzelmann}, {Henri}, {Hermann}, {Hillert},
  {Hinton}, {Hofmann}, {Hofverberg}, {Holler}, {Horns}, {Jacholkowska}, {Jahn},
  {Jamrozy}, {Jung}, {Kastendieck}, {Katarzy{\'n}ski}, {Katz}, {Kaufmann},
  {Kh{\'e}lifi}, {Klochkov}, {Klu{\'z}niak}, {Kneiske}, {Komin}, {Kosack},
  {Kossakowski}, {Krayzel}, {Laffon}, {Lamanna}, {Lenain}, {Lennarz}, {Lohse},
  {Lopatin}, {Lu}, {Marandon}, {Marcowith}, {Masbou}, {Maurin}, {Maxted},
  {Mayer}, {McComb}, {Medina}, {M{\'e}hault}, {Menzler}, {Moderski}, {Mohamed},
  {Moulin}, {Naumann}, {Naumann-Godo}, {de Naurois}, {Nedbal}, {Nguyen},
  {Niemiec}, {Nolan}, {Ohm}, {de O{\~n}a Wilhelmi}, {Opitz}, {Ostrowski},
  {Oya}, {Panter}, {Parsons}, {Paz Arribas}, {Pekeur}, {Pelletier}, {Perez},
  {Petrucci}, {Peyaud}, {Pita}, {P{\"u}hlhofer}, {Punch}, {Quirrenbach},
  {Raue}, {Reimer}, {Reimer}, {Renaud}, {de los Reyes}, {Rieger}, {Ripken},
  {Rob}, {Rosier-Lees}, {Rowell}, {Rudak}, {Rulten}, {Sahakian}, {Sanchez},
  {Santangelo}, {Schlickeiser}, {Schulz}, {Schwanke}, {Schwarzburg},
  {Schwemmer}, {Sheidaei}, {Skilton}, {Sol}, {Spengler}, {Stawarz},
  {Steenkamp}, {Stegmann}, {Stinzing}, {Stycz}, {Sushch}, {Szostek},
  {Tavernet}, {Terrier}, {Tluczykont}, {Valerius}, {van Eldik}, {Vasileiadis},
  {Venter}, {Viana}, {Vincent}, {V{\"o}lk}, {Volpe}, {Vorobiov}, {Vorster},
  {Wagner}, {Ward}, {White}, {Wierzcholska}, {Wouters}, {Zacharias}, {Zajczyk},
  {Zdziarski}, {Zech}, \& {Zechlin}}]{2013A+A...550A...4H}
{H.E.S.S.~Collaboration}, {Abramowski}, A., {Acero}, F., {et~al.}
  2013{\natexlab{b}}, \aap, 550, A4

\bibitem[{{Hinton}(2004)}]{2004NewAR..48..331H}
{Hinton}, J.~A. 2004, New Astronomy Review, 48, 331

\bibitem[{{Impey} \& {Tapia}(1988)}]{1988ApJ...333..666I}
{Impey}, C.~D. \& {Tapia}, S. 1988, \apj, 333, 666

\bibitem[{{Kalberla} {et~al.}(2005){Kalberla}, {Burton}, {Hartmann}, {Arnal},
  {Bajaja}, {Morras}, \& {P{\"o}ppel}}]{2005A+A...440..775K}
{Kalberla}, P.~M.~W., {Burton}, W.~B., {Hartmann}, D., {et~al.} 2005, \aap,
  440, 775

\bibitem[{{Kapahi} {et~al.}(1998){Kapahi}, {Athreya}, {Subrahmanya}, {Baker},
  {Hunstead}, {McCarthy}, \& {van Breugel}}]{1998ApJS..118..327K}
{Kapahi}, V.~K., {Athreya}, R.~M., {Subrahmanya}, C.~R., {et~al.} 1998, \apjs,
  118, 327

\bibitem[{{Kashlinsky}(2005)}]{2005PhR...409..361K}
{Kashlinsky}, A. 2005, \physrep, 409, 361

\bibitem[{{Katarzy{\'n}ski} {et~al.}(2006){Katarzy{\'n}ski}, {Ghisellini},
  {Mastichiadis}, {Tavecchio}, \& {Maraschi}}]{2006A+A...453...47K}
{Katarzy{\'n}ski}, K., {Ghisellini}, G., {Mastichiadis}, A., {Tavecchio}, F.,
  \& {Maraschi}, L. 2006, \aap, 453, 47

\bibitem[{{Katarzy{\'n}ski} {et~al.}(2001){Katarzy{\'n}ski}, {Sol}, \&
  {Kus}}]{2001A+A...367..809K}
{Katarzy{\'n}ski}, K., {Sol}, H., \& {Kus}, A. 2001, \aap, 367, 809

\bibitem[{{Kneiske} \& {Dole}(2010)}]{2010A+A...515A..19K}
{Kneiske}, T.~M. \& {Dole}, H. 2010, \aap, 515, A19

\bibitem[{{Krawczynski} {et~al.}(2004){Krawczynski}, {Hughes}, {Horan},
  {Aharonian}, {Aller}, {Aller}, {Boltwood}, {Buckley}, {Coppi}, {Fossati},
  {G{\"o}tting}, {Holder}, {Horns}, {Kurtanidze}, {Marscher}, {Nikolashvili},
  {Remillard}, {Sadun}, \& {Schr{\"o}der}}]{2004ApJ...601..151K}
{Krawczynski}, H., {Hughes}, S.~B., {Horan}, D., {et~al.} 2004, \apj, 601, 151

\bibitem[{{Lamer} {et~al.}(1996){Lamer}, {Brunner}, \&
  {Staubert}}]{1996A+A...311..384L}
{Lamer}, G., {Brunner}, H., \& {Staubert}, R. 1996, \aap, 311, 384

\bibitem[{{Laurent-Muehleisen} {et~al.}(1998){Laurent-Muehleisen}, {Kollgaard},
  {Ciardullo}, {Feigelson}, {Brinkmann}, \& {Siebert}}]{1998ApJS..118..127L}
{Laurent-Muehleisen}, S.~A., {Kollgaard}, R.~I., {Ciardullo}, R., {et~al.}
  1998, \apjs, 118, 127

\bibitem[{{Laurent-Muehleisen} {et~al.}(1999){Laurent-Muehleisen}, {Kollgaard},
  {Feigelson}, {Brinkmann}, \& {Siebert}}]{1999ApJ...525..127L}
{Laurent-Muehleisen}, S.~A., {Kollgaard}, R.~I., {Feigelson}, E.~D.,
  {Brinkmann}, W., \& {Siebert}, J. 1999, \apj, 525, 127

\bibitem[{{Lenain}(2009)}]{2009PhDT.........5L}
{Lenain}, J.-P. 2009, PhD thesis, LUTH, Observatoire de Paris, CNRS,
  Universit{\'e} Paris Diderot;
  \href{http://tel.archives-ouvertes.fr/tel-00431288}{http://tel.archives-ouve%
rtes.fr/tel-00431288}

\bibitem[{{Li} \& {Ma}(1983)}]{1983ApJ...272..317L}
{Li}, T.-P. \& {Ma}, Y.-Q. 1983, \apj, 272, 317

\bibitem[{{MAGIC Collaboration} {et~al.}(2008){MAGIC Collaboration}, {Albert},
  {Aliu}, {Anderhub}, {Antonelli}, {Antoranz}, {Backes}, {Baixeras}, {Barrio},
  {Bartko}, {Bastieri}, {Becker}, {Bednarek}, {Berger}, {Bernardini},
  {Bigongiari}, {Biland}, {Bock}, {Bonnoli}, {Bordas}, {Bosch-Ramon}, {Bretz},
  {Britvitch}, {Camara}, {Carmona}, {Chilingarian}, {Commichau}, {Contreras},
  {Cortina}, {Costado}, {Covino}, {Curtef}, {Dazzi}, {De Angelis}, {Cea del
  Pozo}, {de los Reyes}, {De Lotto}, {De Maria}, {De Sabata}, {Mendez},
  {Dominguez}, {Dorner}, {Doro}, {Errando}, {Fagiolini}, {Ferenc},
  {Fern{\'a}ndez}, {Firpo}, {Fonseca}, {Font}, {Galante}, {L{\'o}pez},
  {Garczarczyk}, {Gaug}, {Goebel}, {Hayashida}, {Herrero}, {H{\"o}hne}, {Hose},
  {Hsu}, {Huber}, {Jogler}, {Kneiske}, {Kranich}, {La Barbera}, {Laille},
  {Leonardo}, {Lindfors}, {Lombardi}, {Longo}, {L{\'o}pez}, {Lorenz},
  {Majumdar}, {Maneva}, {Mankuzhiyil}, {Mannheim}, {Maraschi}, {Mariotti},
  {Mart{\'{\i}}nez}, {Mazin}, {Meucci}, {Meyer}, {Miranda}, {Mirzoyan},
  {Mizobuchi}, {Moles}, {Moralejo}, {Nieto}, {Nilsson}, {Ninkovic}, {Otte},
  {Oya}, {Panniello}, {Paoletti}, {Paredes}, {Pasanen}, {Pascoli}, {Pauss},
  {Pegna}, {Perez-Torres}, {Persic}, {Peruzzo}, {Piccioli}, {Prada},
  {Prandini}, {Puchades}, {Raymers}, {Rhode}, {Rib{\'o}}, {Rico}, {Rissi},
  {Robert}, {R{\"u}gamer}, {Saggion}, {Saito}, {Salvati}, {Sanchez-Conde},
  {Sartori}, {Satalecka}, {Scalzotto}, {Scapin}, {Schmitt}, {Schweizer},
  {Shayduk}, {Shinozaki}, {Shore}, {Sidro}, {Sierpowska-Bartosik},
  {Sillanp{\"a}{\"a}}, {Sobczynska}, {Spanier}, {Stamerra}, {Stark}, {Takalo},
  {Tavecchio}, {Temnikov}, {Tescaro}, {Teshima}, {Tluczykont}, {Torres},
  {Turini}, {Vankov}, {Venturini}, {Vitale}, {Wagner}, {Wittek}, {Zabalza},
  {Zandanel}, {Zanin}, \& {Zapatero}}]{2008Sci...320.1752M}
{MAGIC Collaboration}, {Albert}, J., {Aliu}, E., {et~al.} 2008, Science, 320,
  1752

\bibitem[{{Malkov} \& {O'C Drury}(2001)}]{2001RPPh...64..429M}
{Malkov}, M.~A. \& {O'C Drury}, L. 2001, Reports on Progress in Physics, 64,
  429

\bibitem[{{Mannheim} {et~al.}(1991){Mannheim}, {Biermann}, \&
  {Kruells}}]{1991A+A...251..723M}
{Mannheim}, K., {Biermann}, P.~L., \& {Kruells}, W.~M. 1991, \aap, 251, 723

\bibitem[{{Mannucci} {et~al.}(2001){Mannucci}, {Basile}, {Poggianti},
  {Cimatti}, {Daddi}, {Pozzetti}, \& {Vanzi}}]{2001MNRAS.326..745M}
{Mannucci}, F., {Basile}, F., {Poggianti}, B.~M., {et~al.} 2001, \mnras, 326,
  745

\bibitem[{{Mattox} {et~al.}(1996){Mattox}, {Bertsch}, {Chiang}, {Dingus},
  {Digel}, {Esposito}, {Fierro}, {Hartman}, {Hunter}, {Kanbach}, {Kniffen},
  {Lin}, {Macomb}, {Mayer-Hasselwander}, {Michelson}, {von Montigny},
  {Mukherjee}, {Nolan}, {Ramanamurthy}, {Schneid}, {Sreekumar}, {Thompson}, \&
  {Willis}}]{1996ApJ...461..396M}
{Mattox}, J.~R., {Bertsch}, D.~L., {Chiang}, J., {et~al.} 1996, \apj, 461, 396

\bibitem[{{Meyer} {et~al.}(2012){Meyer}, {Raue}, {Mazin}, \&
  {Horns}}]{2012A+A...542A..59M}
{Meyer}, M., {Raue}, M., {Mazin}, D., \& {Horns}, D. 2012, \aap, 542, A59

\bibitem[{{M{\"u}cke} \& {Protheroe}(2001)}]{2001APh....15..121M}
{M{\"u}cke}, A. \& {Protheroe}, R.~J. 2001, Astroparticle Physics, 15, 121

\bibitem[{{Murphy} {et~al.}(2010){Murphy}, {Sadler}, {Ekers}, {Massardi},
  {Hancock}, {Mahony}, {Ricci}, {Burke-Spolaor}, {Calabretta}, {Chhetri}, {de
  Zotti}, {Edwards}, {Ekers}, {Jackson}, {Kesteven}, {Lindley}, {Newton-McGee},
  {Phillips}, {Roberts}, {Sault}, {Staveley-Smith}, {Subrahmanyan}, {Walker},
  \& {Wilson}}]{2010MNRAS.402.2403M}
{Murphy}, T., {Sadler}, E.~M., {Ekers}, R.~D., {et~al.} 2010, \mnras, 402, 2403

\bibitem[{{Neronov} {et~al.}(2010){Neronov}, {Semikoz}, \&
  {Vovk}}]{2010ATel.2610....1N}
{Neronov}, A., {Semikoz}, D., \& {Vovk}, I. 2010, The Astronomer's Telegram,
  2610, 1

\bibitem[{{Neronov} {et~al.}(2011){Neronov}, {Semikoz}, \&
  {Vovk}}]{2011A+A...529A..59N}
{Neronov}, A., {Semikoz}, D., \& {Vovk}, I. 2011, \aap, 529, A59+

\bibitem[{{Nolan} {et~al.}(2012){Nolan}, {Abdo}, {Ackermann}, {Ajello},
  {Allafort}, {Antolini}, {Atwood}, {Axelsson}, {Baldini}, {Ballet}, \&
  et~al.}]{2012ApJS..199...31N}
{Nolan}, P.~L., {Abdo}, A.~A., {Ackermann}, M., {et~al.} 2012, \apjs, 199, 31

\bibitem[{{Orr} {et~al.}(2011){Orr}, {Krennrich}, \&
  {Dwek}}]{2011ApJ...733...77O}
{Orr}, M.~R., {Krennrich}, F., \& {Dwek}, E. 2011, \apj, 733, 77

\bibitem[{{Padovani} \& {Giommi}(1995)}]{1995ApJ...444..567P}
{Padovani}, P. \& {Giommi}, P. 1995, \apj, 444, 567

\bibitem[Padovani \& Giommi(1996)]{1996MNRAS.279..526P} Padovani, P., \& Giommi, P.\ 1996, \mnras, 279, 526

\bibitem[{{Pesce} {et~al.}(1995){Pesce}, {Falomo}, \&
  {Treves}}]{1995AJ....110.1554P}
{Pesce}, J.~E., {Falomo}, R., \& {Treves}, A. 1995, \aj, 110, 1554

\bibitem[{{Piron} {et~al.}(2001){Piron}, {Djannati-Atai}, {Punch}, {Tavernet},
  {Barrau}, {Bazer-Bachi}, {Chounet}, {Debiais}, {Degrange}, {Dezalay},
  {Espigat}, {Fabre}, {Fleury}, {Fontaine}, {Goret}, {Gouiffes}, {Khelifi},
  {Malet}, {Masterson}, {Mohanty}, {Nuss}, {Renault}, {Rivoal}, {Rob}, \&
  {Vorobiov}}]{2001A+A...374..895P}
{Piron}, F., {Djannati-Atai}, A., {Punch}, M., {et~al.} 2001, \aap, 374, 895

\bibitem[{{Pita} {et~al.}(2012){Pita}, {Goldoni}, {Boisson}, {Becherini},
  {G{\'e}rard}, {Lenain}, \& {Punch}}]{2012AIPC.1505..566P}
{Pita}, S., {Goldoni}, P., {Boisson}, C., {et~al.} 2012, in American Institute
  of Physics Conference Series, Vol. 1505, American Institute of Physics
  Conference Series, ed. F.~A. {Aharonian}, W.~{Hofmann}, \& F.~M. {Rieger},
  566--569

\bibitem[{{Poole} {et~al.}(2008){Poole}, {Breeveld}, {Page}, {Landsman},
  {Holland}, {Roming}, {Kuin}, {Brown}, {Gronwall}, {Hunsberger}, {Koch},
  {Mason}, {Schady}, {vanden Berk}, {Blustin}, {Boyd}, {Broos}, {Carter},
  {Chester}, {Cucchiara}, {Hancock}, {Huckle}, {Immler}, {Ivanushkina},
  {Kennedy}, {Marshall}, {Morgan}, {Pandey}, {de Pasquale}, {Smith}, \&
  {Still}}]{2008MNRAS.383..627P}
{Poole}, T.~S., {Breeveld}, A.~A., {Page}, M.~J., {et~al.} 2008, \mnras, 383,
  627

\bibitem[{{Roming} {et~al.}(2005){Roming}, {Kennedy}, {Mason}, {Nousek}, {Ahr},
  {Bingham}, {Broos}, {Carter}, {Hancock}, {Huckle}, {Hunsberger}, {Kawakami},
  {Killough}, {Koch}, {McLelland}, {Smith}, {Smith}, {Soto}, {Boyd},
  {Breeveld}, {Holland}, {Ivanushkina}, {Pryzby}, {Still}, \&
  {Stock}}]{2005SSRv..120...95R}
{Roming}, P.~W.~A., {Kennedy}, T.~E., {Mason}, K.~O., {et~al.} 2005, \ssr, 120,
  95

\bibitem[{{Saxton} {et~al.}(2008){Saxton}, {Read}, {Esquej}, {Freyberg},
  {Altieri}, \& {Bermejo}}]{2008A+A...480..611S}
{Saxton}, R.~D., {Read}, A.~M., {Esquej}, P., {et~al.} 2008, \aap, 480, 611

\bibitem[{{Schlegel} {et~al.}(1998){Schlegel}, {Finkbeiner}, \&
  {Davis}}]{1998ApJ...500..525S}
{Schlegel}, D.~J., {Finkbeiner}, D.~P., \& {Davis}, M. 1998, \apj, 500, 525

\bibitem[{{Sikora} \& {Zbyszewska}(1985)}]{1985MNRAS.212..553S}
{Sikora}, M. \& {Zbyszewska}, M. 1985, \mnras, 212, 553

\bibitem[{{Sol} {et~al.}(1989){Sol}, {Pelletier}, \&
  {Asseo}}]{1989MNRAS.237..411S}
{Sol}, H., {Pelletier}, G., \& {Asseo}, E. 1989, \mnras, 237, 411

\bibitem[{{Stecker} {et~al.}(1992){Stecker}, {de Jager}, \&
  {Salamon}}]{1992ApJ...390L..49S}
{Stecker}, F.~W., {de Jager}, O.~C., \& {Salamon}, M.~H. 1992, \apjl, 390, L49

\bibitem[{{Stocke} {et~al.}(1991){Stocke}, {Morris}, {Gioia}, {Maccacaro},
  {Schild}, {Wolter}, {Fleming}, \& {Henry}}]{1991ApJS...76..813S}
{Stocke}, J.~T., {Morris}, S.~L., {Gioia}, I.~M., {et~al.} 1991, \apjs, 76, 813

\bibitem[{{Summerlin} \& {Baring}(2012)}]{2012ApJ...745...63S}
{Summerlin}, E.~J. \& {Baring}, M.~G. 2012, \apj, 745, 63

\bibitem[{{Tanihata} {et~al.}(2004){Tanihata}, {Kataoka}, {Takahashi}, \&
  {Madejski}}]{2004ApJ...601..759T}
{Tanihata}, C., {Kataoka}, J., {Takahashi}, T., \& {Madejski}, G.~M. 2004,
  \apj, 601, 759

\bibitem[{{Tavecchio} {et~al.}(2010{\natexlab{a}}){Tavecchio}, {Ghisellini},
  {Bonnoli}, \& {Ghirlanda}}]{2010MNRAS.405L..94T}
{Tavecchio}, F., {Ghisellini}, G., {Bonnoli}, G., \& {Ghirlanda}, G.
  2010{\natexlab{a}}, \mnras, 405, L94

\bibitem[{{Tavecchio} {et~al.}(2010{\natexlab{b}}){Tavecchio}, {Ghisellini},
  {Ghirlanda}, {Foschini}, \& {Maraschi}}]{2010MNRAS.401.1570T}
{Tavecchio}, F., {Ghisellini}, G., {Ghirlanda}, G., {Foschini}, L., \&
  {Maraschi}, L. 2010{\natexlab{b}}, \mnras, 401, 1570

\bibitem[{{Tavecchio} {et~al.}(1998){Tavecchio}, {Maraschi}, \&
  {Ghisellini}}]{1998ApJ...509..608T}
{Tavecchio}, F., {Maraschi}, L., \& {Ghisellini}, G. 1998, \apj, 509, 608

\bibitem[{{Vaughan} {et~al.}(2003){Vaughan}, {Edelson}, {Warwick}, \&
  {Uttley}}]{2003MNRAS.345.1271V}
{Vaughan}, S., {Edelson}, R., {Warwick}, R.~S., \& {Uttley}, P. 2003, \mnras,
  345, 1271

\bibitem[{{Voges} {et~al.}(1999){Voges}, {Aschenbach}, {Boller},
  {Br{\"a}uninger}, {Briel}, {Burkert}, {Dennerl}, {Englhauser}, {Gruber},
  {Haberl}, {Hartner}, {Hasinger}, {K{\"u}rster}, {Pfeffermann}, {Pietsch},
  {Predehl}, {Rosso}, {Schmitt}, {Tr{\"u}mper}, \&
  {Zimmermann}}]{1999A+A...349..389V}
{Voges}, W., {Aschenbach}, B., {Boller}, T., {et~al.} 1999, \aap, 349, 389

\bibitem[{{Wilks}(1938)}]{1938AnMathStat.9.60W}
{Wilks}, S.~S. 1938, Annals of Mathematical Statistics, 9, 60

\bibitem[{{Wright} \& {Otrupcek}(1990)}]{1990PKS...C......0W}
{Wright}, A. \& {Otrupcek}, R. 1990, in PKS Catalog (1990), 0

\bibitem[{{Wright} {et~al.}(1983){Wright}, {Ables}, \&
  {Allen}}]{1983MNRAS.205..793W}
{Wright}, A.~E., {Ables}, J.~G., \& {Allen}, D.~A. 1983, \mnras, 205, 793

\bibitem[{{Zhang} {et~al.}(1999){Zhang}, {Celotti}, {Treves}, {Chiappetti},
  {Ghisellini}, {Maraschi}, {Pian}, {Tagliaferri}, {Tavecchio}, \&
  {Urry}}]{1999ApJ...527..719Z}
{Zhang}, Y.~H., {Celotti}, A., {Treves}, A., {et~al.} 1999, \apj, 527, 719

\end{thebibliography}

\end{document}